\documentclass{aa}  

\usepackage{graphicx}
\usepackage{txfonts}
\usepackage{multirow}
\usepackage{colortbl}

\usepackage{hyperref}  
\hypersetup{colorlinks=true,linkcolor=[rgb]{1.,0.2,0.2},citecolor=[rgb]{0.1,0.4,1.},filecolor=[rgb]{0.7,0.2,0.2},urlcolor=[rgb]{0.7,0.2,0.2}}

\usepackage{color}
\definecolor{blue}{rgb}{0., 0., 1}
\definecolor{lightblue}{rgb}{0.1,0.4,1.}

\defcitealias{Acebron2022}{A22}
\defcitealias{Sharon2017}{S17}
\begin{document}

   \title{New strong lensing modelling of SDSS J2222+2745 enhanced with VLT/MUSE spectroscopy}


\author{
A.~Acebron \inst{\ref{unimi},\ref{inafmilano}} \fnmsep\thanks{E-mail: \href{mailto:ana.acebron@unimi.it}{ana.acebron@unimi.it}}
\and
C.~Grillo \inst{\ref{unimi},\ref{inafmilano}} \and
P.~Bergamini \inst{\ref{unimi}, \ref{inafbo}} \and
G.~B.~Caminha \inst{\ref{max_plank}} \and
P.~Tozzi \inst{\ref{inaffi}} \and
A.~Mercurio \inst{\ref{inafna}} \and
P.~Rosati \inst{\ref{unife},\ref{inafbo}} \and
G.~Brammer \inst{\ref{dawn},\ref{niels}} \and
M.~Meneghetti \inst{\ref{inafbo}} \and
M.~Nonino \inst{\ref{inafts}} \and
E.~Vanzella \inst{\ref{inafbo}}
}
\institute{
Dipartimento di Fisica, Universit\`a  degli Studi di Milano, via Celoria 16, I-20133 Milano, Italy \label{unimi}
\and
INAF - IASF Milano, via A. Corti 12, I-20133 Milano, Italy \label{inafmilano}
\and
INAF -- OAS, Osservatorio di Astrofisica e Scienza dello Spazio di Bologna, via Gobetti 93/3, I-40129 Bologna, Italy \label{inafbo} 
\and
Max-Planck-Institut f\"ur Astrophysik, Karl-Schwarzschild-Str. 1, D-85748 Garching, Germany \label{max_plank}
\and
INAF – Osservatorio Astrofisico di Arcetri, Largo E. Fermi 5, I-50125, Firenze, Italy \label{inaffi}
\and
INAF -- Osservatorio Astronomico di Capodimonte, Via Moiariello 16, I-80131 Napoli, Italy \label{inafna}
\and
Dipartimento di Fisica e Scienze della Terra, Universit\`a degli Studi di Ferrara, via Saragat 1, I-44122 Ferrara, Italy \label{unife}
\and
Cosmic Dawn Center (DAWN), Copenhagen, Denmark \label{dawn}
\and
Niels Bohr Institute, University of Copenhagen, Jagtvej 128, 2200 Copenhagen, Denmark \label{niels}
\and
INAF -- Osservatorio Astronomico di Trieste, via G. B. Tiepolo 11, I-34143, Trieste, Italy \label{inafts}
           }

   \date{\today}

 
  \abstract
   {SDSS J2222+2745, at $z = 0.489$, is one of the few currently known lens clusters with multiple images (six) of a background ($z = 2.801$) quasar with measured time delays between two image pairs (with a sub-percent relative error for the longer time delay). Systems of this kind can be exploited as alternative cosmological probes through high-precision and accurate strong lensing models.}
   {We present recent observations from the Multi Unit Spectroscopic Explorer (MUSE) on the Very Large Telescope (VLT) and new total mass models of the core of the galaxy cluster SDSS J2222+2745.}
   {We combine archival multi-band, high-resolution imaging from the \textit{Hubble} Space Telescope (HST) with our VLT/MUSE spectroscopic data to securely identify 34 cluster members and 12 multiple images from 3 background sources. 
   We also measure the stellar velocity dispersions of 13 cluster galaxies, down to HST $\rm F160W = 21~mag$, enabling an independent estimate of the contribution of the sub-halo mass component to the lens total mass. 
   By leveraging the new spectroscopic dataset, we build improved strong lensing models.
   }
   {The projected total mass distribution of the lens cluster is best modelled with a single large-scale mass component, a galaxy-scale component, anchored by the VLT/MUSE kinematic information, and an external shear component. The best-fit strong lensing model yields a root mean square separation between the model-predicted and observed positions of the multiple images of $0\arcsec.29$. 
   When analysing the impact of systematic uncertainties, stemming from modelling assumptions and used observables, we find that the resulting projected total mass profile, the relative weight of the sub-halo mass component, and the critical lines are consistent, within the statistical uncertainties. The predicted magnification and time-delay values are, instead, more sensitive to the local details of the lens total mass distribution, and vary significantly among lens models that are similarly good at reproducing the observed multiple image positions. In particular, the model-predicted time delays can differ by a factor of up to $\sim 1.5$.}
   {SDSS J2222+2745 is a promising lens cluster for cosmological applications. However, due to its complex morphology, the relatively low number of secure {`point-like'} multiple images, and current model degeneracies, it becomes clear that additional information (from the observed surface brightness distribution of lensed sources and the measured time delays) needs to be included in the modelling for accurate and precise cosmological measurements. The full VLT/MUSE secure spectroscopic catalogue presented in this work is made publicly available.}


   \maketitle
%

  \begin{figure*}
  \centering
   \resizebox{0.7\hsize}{!}
    {\includegraphics[]{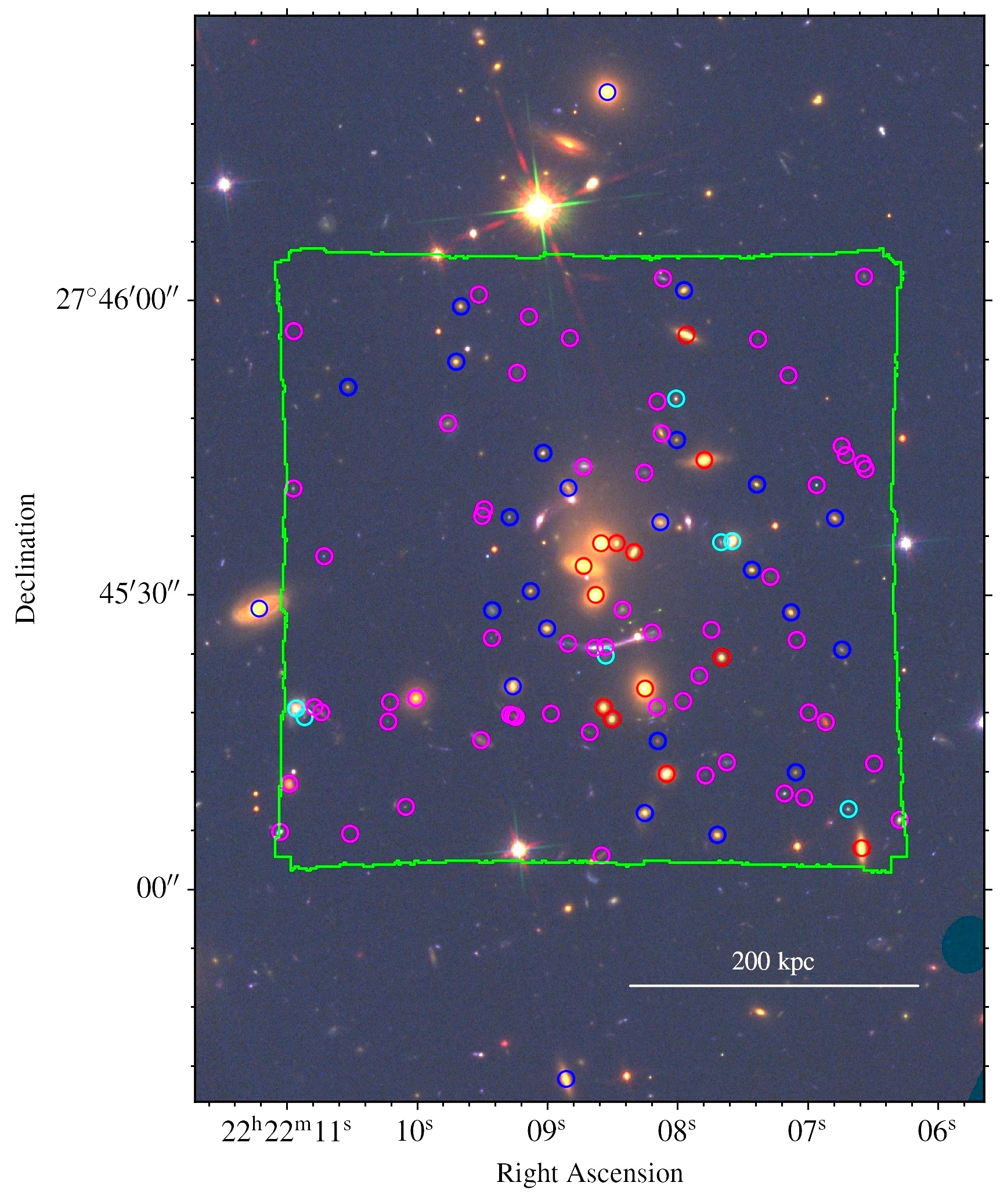}}
    \caption{Colour-composite image of SDSS 2222 obtained by combining the F435W (blue), F814W (green), and F160W (red) HST passbands. The $\rm \sim1~arcmin^2$ MUSE footprint is shown in green. 
    The galaxies with a secure MUSE redshift measurement  ($\mathrm{QF \geq 2}$) are highlighted with colour-coded circles. Cluster galaxies are identified in blue (three cluster members outside the MUSE field of view are included in the SDSS DR9 catalogue); those with a reliable measurement of their stellar velocity dispersion in red; and foreground and background objects in cyan and magenta, respectively. The multiple images are shown in Fig. \ref{Fig:multimg}.}
   \label{Fig:MUSE}
   \end{figure*}

  \begin{figure*}
  \centering
   \resizebox{0.7\hsize}{!}
    {\includegraphics[]{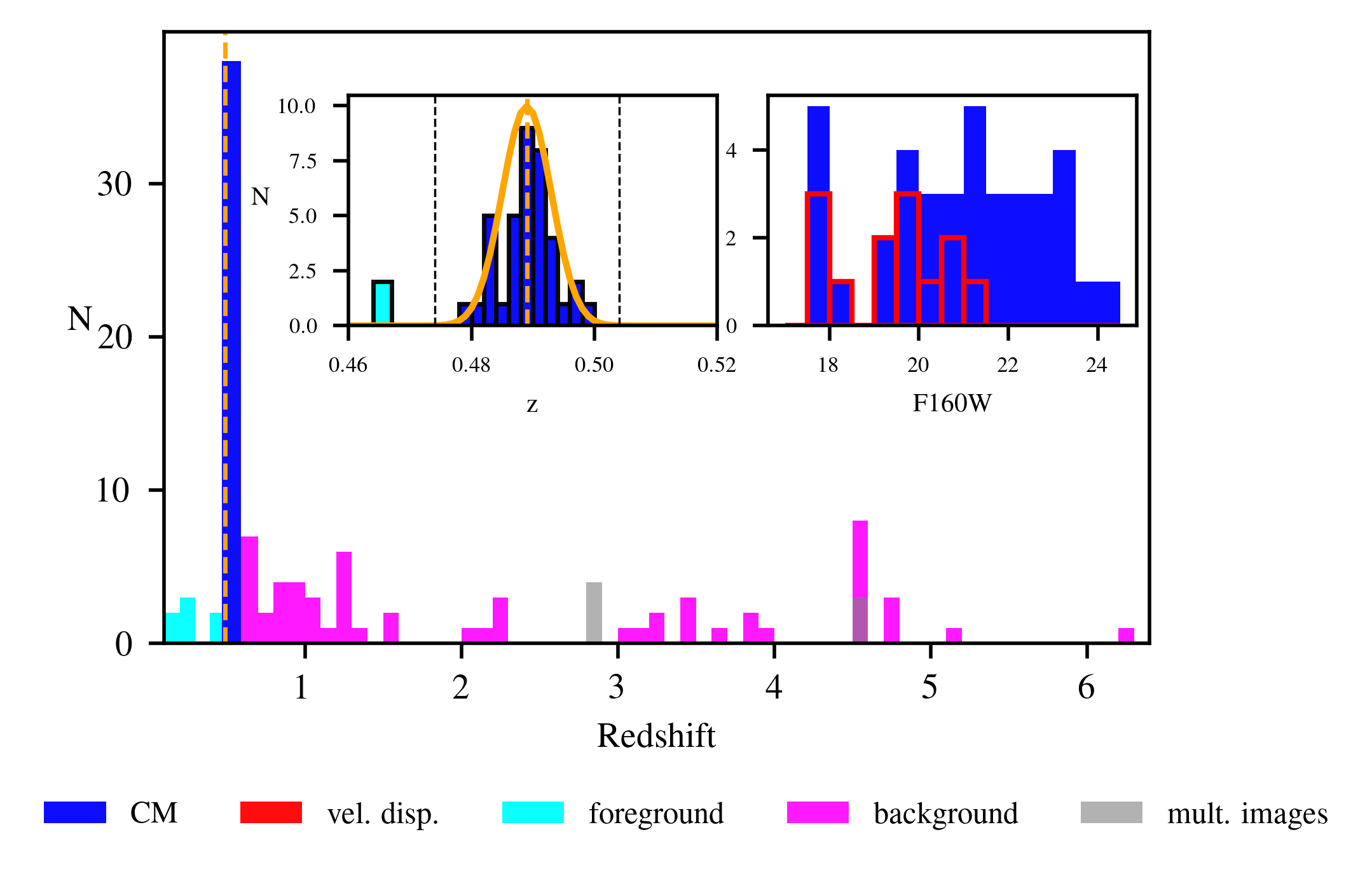}}
    \caption{MUSE spectroscopic redshift distribution of the objects with a $\mathrm{QF \geq 2}$ identified in the SDSS 2222 cluster field. The top-left inset shows a zoomed-in view around the cluster redshift $z = 0.489$, as shown by the mean value of the Gaussian distribution (in orange). The vertical dashed black lines locate the redshift interval [0.474 - 0.504], which includes the 34 spectroscopically confirmed MUSE cluster members. The top-right inset shows the distribution of cluster members as a function of their magnitudes in the F160W HST band. The same colour code as in Fig. \ref{Fig:MUSE} is adopted. The multiple images are marked in grey.}
   \label{Fig:histoMUSE}
   \end{figure*}

\section{Introduction}
Exquisite multi-wavelength observations of galaxy clusters have recently allowed for a new generation of high-precision and accurate strong lensing (SL) models. 
High-resolution \textit{Hubble} Space Telescope (HST) imaging, combined with Multi Unit Spectroscopic Explorer \citep[MUSE;][]{Bacon2010, Bacon2014} spectroscopic follow-up, has enabled the identification of a large number of secure multiple images that critically constrain the SL models \citep[see e.g.][]{Richard2015, Kawamata2016, Caminha2019, Jauzac2019, Lagattuta2019, Bergamini2021}. 
In particular, the inclusion in the models of multiply lensed emission knots within extended sources provides crucial information on the position of the critical lines and therefore on the high-magnification regions \citep{Grillo2016, Bergamini2021, Bergamini2022}. 
In parallel, MUSE observations allow for the secure identification of large sets of cluster galaxies \citep{Mercurio2021, Lagattuta2022}. Their stellar kinematics can be used effectively in SL models to independently weight the contribution of the sub-halo mass component, thus reducing inherent model degeneracies \citep[][]{Bergamini2019, Bergamini2021, Pignataro2021}.
 \citet{Granata2022}, exploiting the MUSE kinematic measurements of cluster members, together with their structural parameters determined from HST photometry, adopted the Fundamental Plane relation \citep{Dressler1987} to obtain a more realistic description of the total mass distribution of cluster galaxies in the galaxy cluster Abell S1063. 

Cluster lenses, such as SDSS J1004+4112 \citep{Inada2003}, SDSS J1029+2623 \citep[][hereafter \citetalias{Acebron2022}]{Oguri2013, Acebron2022}, MACS J1149.5+2223 \citep{Grillo2016}, and SDSS J2222+2745, where a variable background source is multiply imaged, are in addition of particular interest for their cosmological applications \citep[see e.g.][]{Grillo2018, Grillo2020}. They represent emergent, independent probes for measuring the expansion rate \citep{Refsdal1964} and the geometry of the Universe. In the era of precision cosmology, these cluster systems can offer important insights into unknown systematic effects and help clarify current tensions in cosmology \citep[see][for a recent review]{Moresco2022}. 
   
SDSS J2222+2745 (SDSS 2222, hereafter), at a redshift of $z=0.489$, was discovered within the Sloan Giant Arcs Survey \citep{Hennawi2008, Bayliss2011, Sharon2020} from the Sloan Digital Sky Survey (SDSS) Data Release 8 \citep[DR8;][]{Aihara2011}.
Subsequent photometric and spectroscopic follow-up, with the MOsaic CAmera (MOSCA) and the Andalusia Faint Object Spectrograph and Camera (ALFOSC) at the Nordic Optical Telescope (NOT), revealed that SDSS 2222 is a lens cluster that produces six multiple images of a background quasi-stellar object \citep[quasar or QSO;][]{Dahle2013}. These observations provided a spectroscopic confirmation of the six images of the QSO (labelled A, B, C, D, E, and F) and of the southern arc, which were used to build a first SL model of the galaxy cluster.
Following its discovery, SDSS 2222 was photometrically monitored with the NOT to measure the time delays between the QSO multiple images. Observations taken between September 2012 and January 2019 yielded the time-delay values of $\Delta t_{AB}=-42.44^{+1.44}_{-1.36}$ days and $\Delta t_{AC}=696.65^{+2.10}_{-2.00}$ days \citep{Dahle2015, Dyrland2019}. 
High-resolution imaging obtained with the HST and spectroscopic follow-up data from the Gemini Multi-Object Spectrograph (GMOS) on the Gemini-North Telescope were then used to create an updated SL model in \citet[][hereafter \citetalias{Sharon2017}]{Sharon2017}. Additional multiple image candidates were identified in the HST images, while GMOS spectroscopy refined the redshift estimates for the six quasar images and the southern arc and yielded new redshift measurements for one multiple image and 11 cluster galaxies. 

In this work we present recent spectroscopic observations of SDSS 2222 obtained with the MUSE integral field spectrograph, mounted on the Very Large Telescope (VLT). We exploit the newly obtained data to build an improved SL model of the cluster. The paper is organised as follows. Section \ref{sec:data} describes the HST imaging and the VLT/MUSE data used to develop the new lens model of SDSS 2222. Section \ref{sec:SLM} presents the selection of the multiple images and the cluster members, as well as the adopted total mass parametrisation of the cluster. We discuss our results in Sect. \ref{sec:results} and draw the main conclusions of this work in Sect. \ref{sec:conclusions}.

Throughout this work we adopt a flat $\Lambda$ cold dark matter cosmology with $\Omega_{\rm m} = 0.3$ and $H_0= 70\,\mathrm{km\,s^{-1}\,Mpc^{-1}}$. Within this cosmology, a projected distance of $1\arcsec$ corresponds to a physical scale of 6.03 kpc at the cluster redshift of $z=0.489$. All magnitudes are given in the AB system \citep{Oke1974}. The quoted uncertainties correspond to the $68\%$ confidence interval, unless otherwise stated.

  \begin{figure*}
  \centering
    \resizebox{\hsize}{!}
    {\includegraphics[]{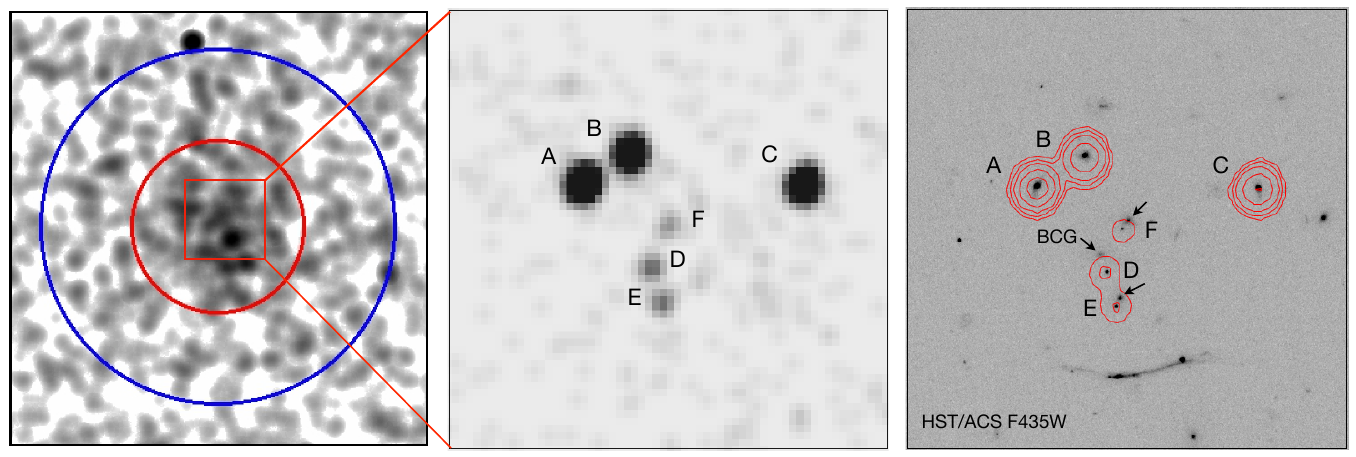}}
    \caption{{\sl Chandra} observations of SDSS 2222. \textit{Left panel}: {\sl Chandra} image in the soft band (0.5-2.0 keV) of a region of $2.4\times 2.4$ arcmin$^2$ centred on the BCG of SDSS 2222. 
    The image has been heavily smoothed with a kernel of $6\arcsec$ following the removal of the six X-ray images of the lensed quasar, readily detected in the 0.5-7 keV band (\textit{central panel}). The red and blue circles correspond to a radius of $33\arcsec$ (200 kpc at the cluster redshift) and $70\arcsec$, respectively. \textit{Right panel}:  HST/ACS F435W image of the cluster core (30\arcsec\ across) with {\sl Chandra}  contours overlaid in red (0.5-7 keV), showing the six QSO images centred on the peak of the X-ray sources. The three central passive galaxies are relatively faint in the F435W band (arrows).}
   \label{Fig:Xray}
   \end{figure*}
      
\section{Data} \label{sec:data}

We present here the data exploited to develop a new SL model of SDSS 2222: the high-resolution imaging in Sect. \ref{sec:hst} and the newly obtained VLT/MUSE spectroscopic follow-up in Sect. \ref{sec:vlt}. In Sect. \ref{sec:chandra} we briefly present a new analysis of the {\sl Chandra} X-ray observations, although they are not used in the subsequent analysis.

\subsection{HST imaging} \label{sec:hst}

We used archival HST multi-colour imaging (GO-13337; P.I.: Sharon), from the {Advanced Camera for Surveys} (ACS) and the {Wide Field Camera 3} (WFC3), taken in August and October 2014. SDSS 2222 was imaged over two orbits in each of the ACS filters (F475W, F606W, and F814W), while the WFC3 imaging was allocated one single orbit (in F110W and F160W). A detailed description of the observations and data reduction process is provided in \citetalias{Sharon2017}. 
We extracted our HST/F160W photometric catalogue with the public software \texttt{SourceExtractor} \citep{Bertin1996}. 

\subsection{VLT/MUSE spectroscopy} \label{sec:vlt}

SDSS 2222 was targeted with the integral field spectrograph MUSE at the VLT, under programme 0103.A-0554(A) (P.I.: Grillo) between 2019 June 3 and July 10. 
The lens galaxy cluster was observed with a single pointing ($\sim$~1~arcmin$^2$ field of view; see the footprint in Fig. \ref{Fig:MUSE}), for a total of 11 exposures of 1440 seconds each, resulting in a cumulative exposure time of $4.4$ hours on target.

Following the procedure described in \citet{Caminha2017, Caminha2017b, Caminha2019}, we used the standard reduction pipeline \citep[version 2.6;][]{Weilbacher2020} to process the raw MUSE exposures and create the final stacked datacube.
In addition, the `{auto-calibration'} method and the Zurich Atmosphere Purge \citep[ZAP;][]{Soto2016} were applied to mitigate slice-to-slice flux variations and improve the sky-subtraction.
The data have a spatial pixel size of $0\arcsec.2$ and the value of the full width half maximum (FWHM) measured from the white image is $0\arcsec.8$.
Redshifts were then measured by extracting one-dimensional spectra of all sources with HST detections within circular apertures of $0.\arcsec8$ radius. To improve the signal to noise ratio ($S/N$) of the extracted spectra for faint galaxies, we adopted customised apertures based on their estimated morphology from the HST imaging. 
Following \citet{Balestra2016} and \citet{Caminha2016}, we assigned a quality flag (QF) to each redshift measurement in order to quantify its reliability: `{insecure'} (QF = 1), `{likely'} (QF = 2), `{secure'} (QF = 3), or `{based on a single emission line'} (QF = 9).

The full MUSE spectroscopic catalogue contains 118 reliable (i.e. QF $\ge$ 2) redshift measurements, of which 11 are stars, 7 are foreground galaxies ($z < 0.474$), 34 are cluster members ($0.474\le z \le 0.504$; see Sect. \ref{sec:cm}), 59 are background galaxies ($z>0.504$), and 7 are multiple images (see Fig. \ref{Fig:histoMUSE}).
The foreground and background objects are identified in Fig. \ref{Fig:MUSE}, and their coordinates and redshifts are listed in Table \ref{tab:fgbg}. The multiple image and cluster member catalogues are presented in Sects. \ref{sec:multim} and \ref{sec:cm}, respectively.

\subsection{X-ray data} \label{sec:chandra}

SDSS 2222 was observed by the {\sl Chandra} telescope between 2016 April 24 and 29 with the Advanced CCD Imaging Spectrometer (ACIS-S; Observation IDs 17048, 18831, and 18832; P.I.: Pooley), for a total exposure time of 66.06 ks after data reduction. 
Thanks to the exquisite arcsecond resolution of {\sl Chandra}, the six multiple images of the quasar are clearly detected (unblended) and can therefore be removed. A visual inspection of the image allowed us to identify a diffuse emission with a very low surface brightness, and no significant concentration towards the cluster centre, that is associated with the central position of the brightest cluster galaxy (BCG) identified in the HST images. 
This is shown in the left panel of Fig. \ref{Fig:Xray}, where the soft-band diffuse emission is clearly seen inside a radius of $33\arcsec$ (or 200 kpc at the cluster redshift; red circle). However, some emission may also be present out to a radius of $70\arcsec$ (or 420 kpc at the cluster redshift; blue circle). 

Aperture photometry provides $217\pm 22$ and $73 \pm 22$ net counts in the soft and hard band, respectively, within a radius of $33\arcsec$, which maximises the $S/N$ of the diffuse emission in the total band (0.5-7.0 keV). This corresponds to a flux of $1.9\times 10^{-14}$ and $1.7\times 10^{-14}$ ergs~cm$^{-2}$~s$^{-1}$ in the 0.5-2.0 and 2.0-10.0 keV bands, respectively.
The presence of an emission beyond $33\arcsec$ is confirmed by the values of $359\pm 38$ and $135 \pm 42$ net counts in the soft and hard band, respectively, measured within a radius of $70\arcsec$.  These results critically depend on the background subtraction, which was performed by sampling the source-free regions beyond that radius. We verified that the positive photometry obtained in both bands, within $70\arcsec$, is robust against the uncertainty on the soft and hard background. We also note that masking out the quasar images implies a loss of less than 2\% of the solid angle with respect to a full circle with a radius of $33\arcsec$ and therefore does not affect our measurements. 

We find that about 65\% of the signal within $33\arcsec$ in the soft band is contributed by the six quasar images. 
A spectral analysis of the diffuse emission shows that a single-temperature {\tt mekal} model provides a best-fit temperature of $kT=3.5_{-0.8}^{+1.2}$ keV. 
The luminosity of this diffuse emission, corrected for Galactic absorption, is  $\sim\! 2 \times 10^{43}$ ergs~s$^{-1}$ within $30\arcsec$ in the 0.5-2.0 band ($\sim\! 3 \times 10^{43}$ ergs~s$^{-1}$ within $70\arcsec$).
We note, however, that, due to the low S/N, the thermal model is statistically equivalent to a power law with slope $\Gamma = 2.2$.  
We conclude that, with the current X-ray data, it is not possible to test the thermal nature of the diffuse emission and apply the hydrostatic equilibrium to derive a robust X-ray mass profile. Therefore, even though a diffuse intracluster medium component is clearly detected, we do not use the X-ray data to further constrain the virial mass of SDSS 2222 in the following analysis.

\begin{table}
\caption{Coordinates and spectroscopic redshifts, with the corresponding quality flag, of the multiple image systems.} 
\label{tab:multim}
\centering
\begin{tabular}{cccccc}
\hline\hline
ID & R.A. & Decl & $\rm z_{spec}$\tablefootmark{a} & QF & MUSE ID \\ 
& deg & deg &  &  &  \\ 
\hline
A & 335.537698 & 27.760538 & 2.801 & 3 & 1679\\ 
B & 335.536677 & 27.761115 & 2.801 & 3 & 1844\\ 
C & 335.532955 & 27.760503 & 2.801 & 3 & 1789\\ 
D & 335.536191 & 27.758897 & 2.801 & 2 & 1885\\ 
E & 335.535998 & 27.758252 & 2.801 & - & - \\ 
F & 335.535866 & 27.759723 & 2.801 & - & - \\ 
1a & 335.537495 & 27.760796 & 2.801 & - & - \\ 
1b & 335.536859 & 27.761150 & 2.801 & - & - \\ 
1c & 335.532917 & 27.760339 & 2.801 & - & - \\ 
2a & 335.538397 & 27.758230 & 4.560 & 3 & 392 \\ 
2b & 335.534815 & 27.757629 & 4.560 & 3 & 435 \\ 
2c & 335.533866 & 27.757973 & 4.560 & 3 & 398 \\ 
3a & 335.538412 & 27.760379 & - & - & - \\ 
3b & 335.535499 & 27.761519 & - & - & - \\ 
3c & 335.533613 & 27.760871 & - & - & - \\ 
4a & 335.534535 & 27.754882 & 4.505 & 1 & 438 \\ 
4b & 335.534083 & 27.754935 & 4.505 & 1 & 2153 \\ 
4c & 335.533520 & 27.755171 & 4.505 & 1 & 555555\\ 
\hline        
\end{tabular}
\tablefoot{
\tablefoottext{a}{Redshift values within a system are averages of the measurements with the highest QFs.}
}
\end{table}

\section{Strong lensing modelling} \label{sec:SLM}

We modelled the total mass distribution of SDSS 2222 with the public software \texttt{lenstool}\footnote{\url{https://projets.lam.fr/projects/lenstool}} \citep[see][for a detailed description]{Kneib1996, Jullo2007}. The SL modelling methodology closely follows that presented in \citetalias{Acebron2022} for another galaxy cluster with multiple images of a background QSO, SDSS J1029+2623. We thus refer the reader to that publication for an in-depth description of the adopted methodology.

In Sect. \ref{sec:multim} we present the multiple images that are used as constraints in the SL model. Section \ref{sec:cm} describes the spectroscopic and photometric selection and the measurement of the stellar velocity dispersion of the cluster members. The general modelling formalism is discussed in Sect. \ref{sec:method} and our adopted parametrisation of SDSS 2222 in Sect. \ref{sec:MassModel}.

\subsection{Multiple image systems}  \label{sec:multim}
The multiple images identified in SDSS 2222, whose positions and redshifts constitute the constraints for our lensing model, were identified thanks to MOSCA observations at the NOT \citep{Dahle2013} and high-resolution HST imaging (\citetalias{Sharon2017}). In this work we adopt the same notation as that introduced in the literature for the QSO system (\cite{Dahle2013}, \citetalias{Sharon2017}) but choose to use  the \texttt{lenstool} convention for the other systems (i.e. a number and a letter identifying, respectively, the family and the image within the family). All systems were analysed using the new MUSE datacube and are discussed below.

\begin{figure*}
  \centering
   \resizebox{0.7\hsize}{!}
    {\includegraphics[]{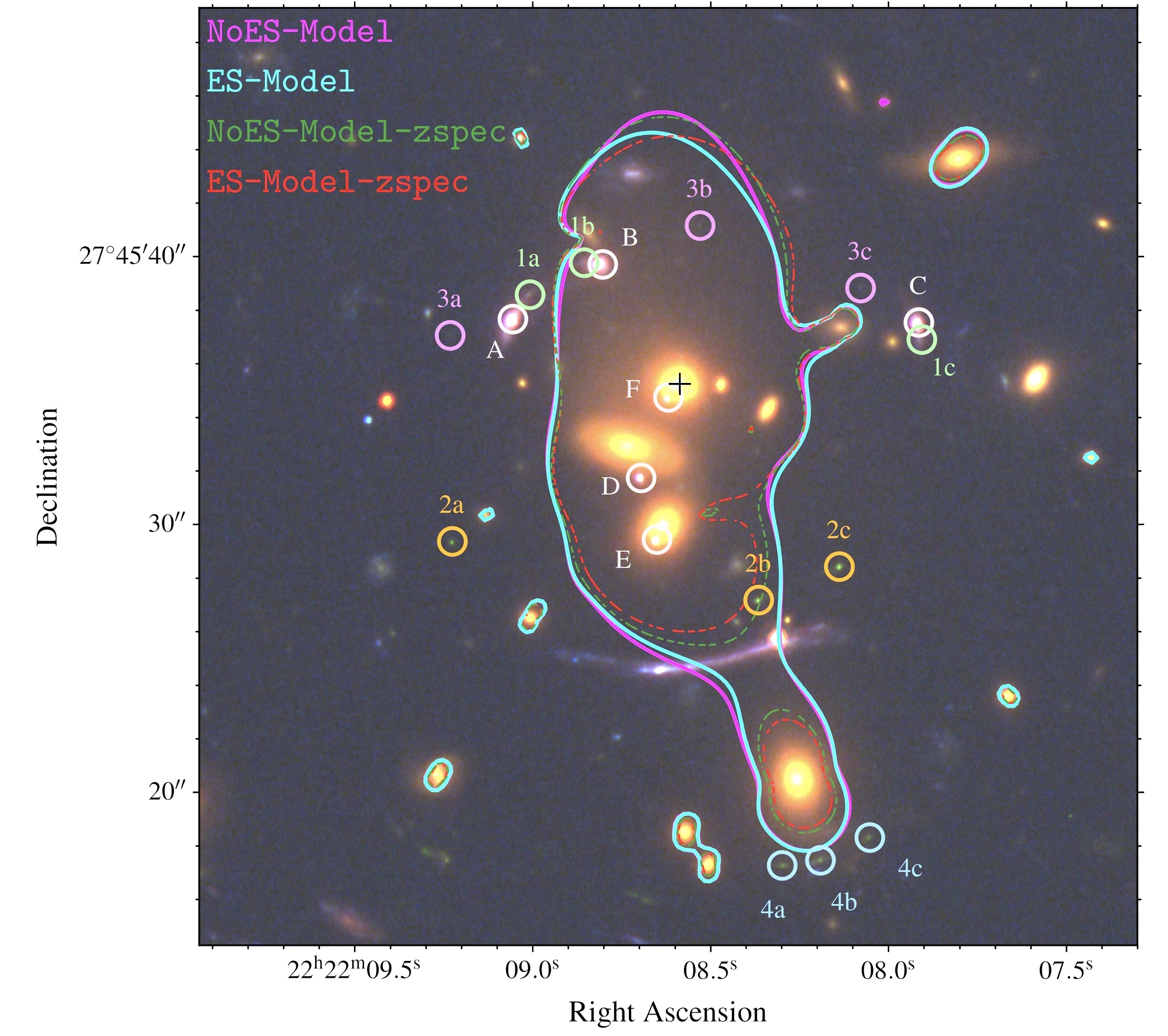}}
    \caption{Red-green-blue image of the central region of SDSS 2222, showing the multiply imaged sources that have been identified (see also Table \ref{tab:multim}).
    We overlay the best-fit critical curves at the quasar redshift ($z=2.801$) from the models presented in this work. The solid and dashed lines display the critical curves for the associated models considering, respectively, the full and spectroscopic-only multiple image samples. The black cross indicates the reference position, corresponding to the luminosity centroid of the BCG (ID 2010), from which the cumulative projected total mass profiles are computed.}
   \label{Fig:multimg}
   \end{figure*}

The background QSO is multiply lensed into six images, labelled A, B, C, D, E, and F (see Fig. \ref{Fig:multimg}). The multiple images were first identified with NOT/MOSCA observations in \citet{Dahle2013}. NOT/ALFOSC follow-up provided a spectroscopic confirmation for images A, B, C, D, and E, while a tentative Ly-$\alpha$ emission line was detected for image F \citep{Dahle2013}. GMOS data spectroscopically confirmed the lensed nature of images A, B, C, and D (\citetalias{Sharon2017}). Based on our MUSE spectra, we refined the spectroscopic redshift value of this system to $z=2.801$ (see Table \ref{tab:multim}). Due to the strong light contamination from bright cluster members, a redshift measurement for images E and F could not be obtained.

System 1 is associated with a knot identified in the galaxy that hosts the QSO and is detected three times in the HST images (close to the most magnified images of the QSO, images A, B, and C). Its redshift value was chosen to be equal to that of the QSO.

The southern arc, labelled A1 in \citetalias{Sharon2017}, is securely confirmed to be at a redshift of $z=2.295$ (see Table \ref{tab:fgbg}). While the extended image clearly shows several emission regions, it is not possible to robustly identify counter-images (misidentifications would in turn introduce potential biases into the lens model). For that reason, this system is not used here. The extended surface brightness modelling of the arc could provide useful constraints on the total mass distribution of the cluster \citep[see][for instance]{Wang2022}. We defer that analysis to a future work.

System 2 is composed of three multiple images, only one of which had a spectroscopic confirmation until now (see system B in \citetalias{Sharon2017}). Thanks to the MUSE observations, all three images are now securely identified at a redshift of $z=4.560$ (see Table \ref{tab:multim}).

System 3 (labelled C in \citetalias{Sharon2017}) is formed by three faint multiple images in the northern region of the cluster. No secure redshift estimate was possible based on the GMOS data (\citetalias{Sharon2017}), and no emission line was detected in the MUSE data.

For the three images of system 4, the GMOS data provided an uncertain measurement of $z=0.86$, possibly contaminated by a foreground object (see system D in \citetalias{Sharon2017}). The analysis of the MUSE datacube results in a tentative (QF = 1) redshift estimate of $z \sim 4.51$, based on a possible Ly-$\alpha$ emission (at the noise level).

The sample of secure multiple image systems spans the redshift range $z=2.801$ to $z=4.560$, with a total of 12 multiple images from three background sources. Systems 3 and 4, which add six additional multiple images, are instead considered photometric, and their redshift values were optimised in the modelling, when included.
We measured the coordinates of the luminosity peaks of the multiple images in the HST F606W band and used them as observables in the lens models.
We show the measured positions of the 18 multiple images in Fig. \ref{Fig:multimg}, and their properties are summarised in Table \ref{tab:multim}. Finally, Fig. \ref{fig:spec} presents the extracted spectra for the images with a QF $\geq$ 2, together with small cutouts of the HST colour-composite image.

  \begin{figure}
  \centering
   \resizebox{\hsize}{!}
    {\includegraphics[]{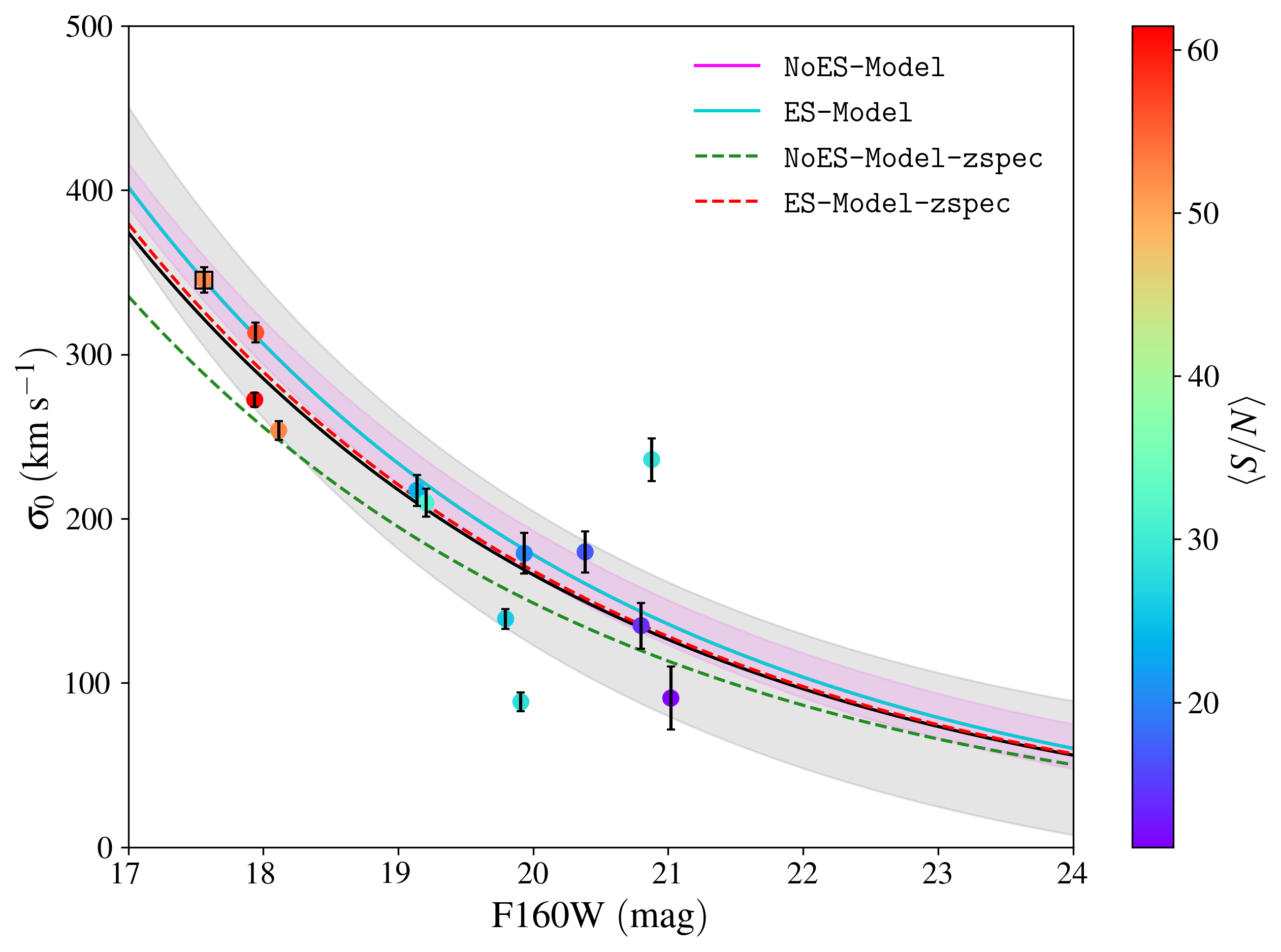}}
    \caption{Measured stellar velocity dispersions for a sub-sample of 13 cluster members as a function of their magnitudes in the HST F160W band (filled circles, colour-coded according to their spectral $\langle S/N\rangle$). The black square pinpoints the values of the BCG. The solid black line and shaded area correspond to the best-fit $\sigma_0$--F160W relation and the 68\% confidence interval, respectively (see Sect. \ref{sec:cm}). 
    We note that all the lens models assume the same value of the slope, $\alpha$, and only the value of the scaling-relation normalisation is optimised within a Gaussian prior centred on the best-fit value of $\sigma_0$. 
    The solid magenta and cyan lines represent the median $\sigma_0$--F160W relations for the \texttt{NoES-Model} and \texttt{ES-Model}, respectively. The resulting relations from the two lens models are consistent and do completely overlap. The dashed lines show the median relations for the corresponding SL models that only include the spectroscopic sample of multiple images. For visual clarity, we only show the $1\sigma$ statistical uncertainties for the \texttt{NoES-Model} (magenta shaded area).}
   \label{Fig:sigma}
   \end{figure}

\subsection{Selection and internal kinematics of cluster members} \label{sec:cm}
The cluster member catalogue was built by considering only secure cluster galaxies. Spectroscopic cluster members were selected mainly based on the analysis of the MUSE datacube. 
From the MUSE data, we identified 34 member galaxies in the cluster inner region with a reliable redshift estimate (i.e. with a QF $\geq$ 2), 26 of which are newly spectroscopically confirmed.
We also included three cluster galaxies outside the MUSE field of view that had previous archival redshift measurements from the SDSS DR9 catalogue (see Fig. \ref{Fig:MUSE} and Table \ref{tab:cm}).
The redshift distribution of these galaxies, shown in Fig. \ref{Fig:histoMUSE}, can be fit with a Gaussian distribution with mean and standard deviation values of $\overline{z}=0.489 \pm 0.004$. 
These spectroscopic cluster members are selected as the galaxies with rest-frame relative velocities within $\Delta V = 3000 ~ \rm km\,s^{-1}$ of the cluster mean velocity, corresponding to the redshift range $z=0.474$-$0.504$.
Their properties are summarised in Table \ref{tab:cm}. The redshift and F160W magnitude distributions of the member galaxies are shown in Fig. \ref{Fig:histoMUSE}.

In a second step, we took further advantage of the MUSE data to measure the line-of-sight stellar velocity dispersion for
the brightest cluster members. Including additional, independent information, such as stellar kinematics, in SL models can help reduce degeneracies between the different cluster mass components. We followed the methodology presented in \citet{Bergamini2019, Bergamini2021}, which has already been applied to several other lens clusters \citep[][]{Pignataro2021, Granata2022, Acebron2022, Bergamini2022}. 
Spectra of the confirmed cluster members were extracted from the MUSE datacube within $0\arcsec.8$ radius apertures and are consistent with the median FWHM value of the MUSE observations (see Sect. \ref{sec:vlt}). We performed a visual inspection and verified, using a small sub-set of galaxies with angularly close neighbours, that reducing the extraction apertures to radii with values of $0\arcsec.6$ yields consistent measurements, given the uncertainties. We eventually only decreased the value of the aperture radius to $0\arcsec.4$ for the BCG (ID 2010 in Table \ref{tab:cm}) due to the contamination from the light of multiple image F of the QSO and of an angularly close satellite galaxy. 
We then measured the line-of-sight stellar velocity dispersion values with the public penalised pixel-fitting method software pPXF \citep{Cappellari2004, Cappellari2017}. The cross-correlation between the observed spectra and an extended set of stellar templates was performed in the rest-frame wavelength range [3600, 5000] \r{A}. 
In order to only exploit reliable measurements in the subsequent lensing analysis, we limited the sample to galaxies with $\langle S/N \rangle >10$ and $\sigma_0 > 80~ \mathrm{km~s^{-1}}$ \citep[as discussed in][]{Bergamini2019, Bergamini2021}. 
Of the 34 MUSE cluster galaxies, 13 of them, down to $ \mathrm{F160W} \sim 21$, satisfy these criteria. 
The measured stellar velocity dispersion values of the 13 cluster members were then fitted following the Bayesian approach presented in \citet{Bergamini2019, Bergamini2021}.
In this way, we derived the best-fit values of the logarithmic slope, $\alpha$, and of the reference velocity dispersion, $\sigma_{0}^{\star}$, of the $\sigma_0$--F160W relation, which were then adopted as prior information for the scaling relations in our SL model of SDSS 2222 (see Sect. \ref{sec:MassModel}). 
We show in Fig. \ref{Fig:sigma} the measured stellar velocity dispersion as a function of the F160W magnitude values for the 13 selected cluster galaxies and the resulting best-fit relation. 

\begin{table}
\caption{Figure-of-merit estimators for the spectroscopic and photometric (top) and the spectroscopic-only (bottom) best-fit models.} 
\label{tab:FoM}
\centering
\begin{tabular}{|c|ccccc|}
\hline
Model & rms [\arcsec]& $\nu$ & $\chi^2$ & BIC & AIC \\
\hline
\texttt{NoES-Model} & 0.34 & 16 & 32.70 & 71.3 & 39.7 \\
\texttt{ES-Model} & 0.29 & 14 & 24.60 & 69.8 & 35.0 \\
\hline       
\texttt{NoES-Model-zspec} & 0.21 & 10 & 8.27 & 30.3 & 13.8 \\
\texttt{ES-Model-zspec} & 0.20 & 8 & 7.98 & 36.4 & 17.5 \\
\hline
\end{tabular}
\tablefoot{
We note that these values correspond to the models before the re-scaling of the multiple image positional uncertainty (see Sect. \ref{sec:MassModel}).
 }
\end{table}

\subsection{Modelling methodology} \label{sec:method}
The pipeline \texttt{lenstool} allows for parametric mass reconstructions of a lens, where the total mass of the lens can be separated into several components. In this work we consider cluster-scale and galaxy-scale mass components.

We chose to model all halos with dual pseudo-isothermal elliptical mass density (dPIE) profiles \citep{Eliasdottir2007}. There are seven free parameters associated with the dPIE profile in \texttt{lenstool}: the coordinates of the centre, $x,~y$;  the ellipticity, $e=(a^2-b^2)/(a^2+b^2)$, where $a$ and $b$ are the values of the major and minor semi-axes, respectively; the orientation, $\theta$ (counted anti-clockwise from the $x$-axis); the core and truncation radii, $r_{\rm core}$ and $r_{\rm cut}$; and a velocity dispersion, $\sigma_{\rm LT}$, which is linked to the central velocity dispersion of the dPIE profile according to the relation $\sigma_0 = \sqrt{3/2}~ \sigma_{\rm LT}$.

Cluster-scale halos are described by non-truncated, elliptical dPIE profiles. Galaxy-scale halos, which are associated with the cluster galaxies, are instead modelled with singular, circular dPIE profiles. To significantly reduce the number of free parameters, the following two scaling relations \citep{Jullo2007} between the galaxy total mass and its corresponding luminosity (as measured in the HST F160W band; see Sect. \ref{sec:cm}) are typically adopted:

\begin{equation}
\label{eqsigma}
\sigma_0=\sigma_0^{\star} \left(\frac{L}{L^{\star}}\right)^{\alpha}
\text{and}~
r_{\rm cut}=r_{\rm cut}^{\star}\left(\frac{L}{L^{\star}}\right)^{\beta},
\end{equation}
where $L^{\star}$ represents the reference luminosity value of a galaxy at the cluster redshift, which we associated with the BCG (with a magnitude value in the HST F160W band of $17.56$). The two free parameters in the lens model are then $\sigma_0^{\star}$ and $r_{\rm cut}^{\star}$. The (fixed) parameters $\alpha$ and $\beta$ correspond to the slopes of the $\sigma_0$ and $r_{\rm cut}$ scaling relations, respectively.
We adopted values of $\alpha$ and $\beta$ so that the galaxy total mass-to-light ratio varies with the luminosity as $M^{\rm tot}_i L_i^{-1} \propto L ^{\gamma}_i$, with $\gamma=0.2$ (i.e. a relation that is compatible with the so-called tilt of the Fundamental Plane; \citealt{Faber1987, Bender1992, Ciotti1996, Bernardi2003, Grillo2010}).

To introduce further flexibility into the lens modelling of galaxy clusters, an external shear component can be considered. The external shear is described by two additional free parameters: its magnitude, $\gamma_{\rm ext}$, and orientation, $\phi_{\rm ext}$.

The best-fitting values of the model parameters that describe the total mass distribution of the lens, $\mathbf{p}$, were obtained by minimising on the image plane the distance between the observed , $\boldsymbol{\theta}^{\rm obs}$, and model-predicted, $\boldsymbol{\theta}^{\rm pred}$, positions of the multiple images through a $\chi^2$ function (see e.g. Eq. 4 in \citetalias[]{Acebron2022}).
More generally, we quote the root mean square (rms) value of the difference between the observed and model-predicted positions of the multiple images to quantify the goodness of our models: 
\begin{equation}
{\rm rms}=\sqrt{\frac{1}{N_{\rm tot}}\sum_{i=1}^{N_{\rm tot}}\left|\boldsymbol{\theta}^{\rm obs}_{i}-{\boldsymbol{\theta}}^{\rm pred}_{i}(\mathbf{p})\right|^2},
\end{equation}
where $N_{\rm tot}$ is the total number of images.

In addition, other statistical estimators can be used to compare lens models with different numbers of observables or free parameters, especially for models with a low number of observable quantities, as in the case of SDSS 2222.
Thus, we also consider in the following the Bayesian information criterion \citep[BIC;][]{Schwarz1978} and the Akaike information criterion \citep[AIC;][]{Akaike1974}.
The BIC is defined as ${\rm BIC}=-2\ln({\mathcal{L}}) + k \times \ln(n) $, and the value of the AIC is obtained as ${\rm AIC}=2k -2\ln({\mathcal{L}})$, where $\mathcal{L}$ is the maximum value of the likelihood, $k$ is the number of free parameters, and $n$ is the number of observational constraints.

\begin{figure}
\centering
\includegraphics[width=\columnwidth]{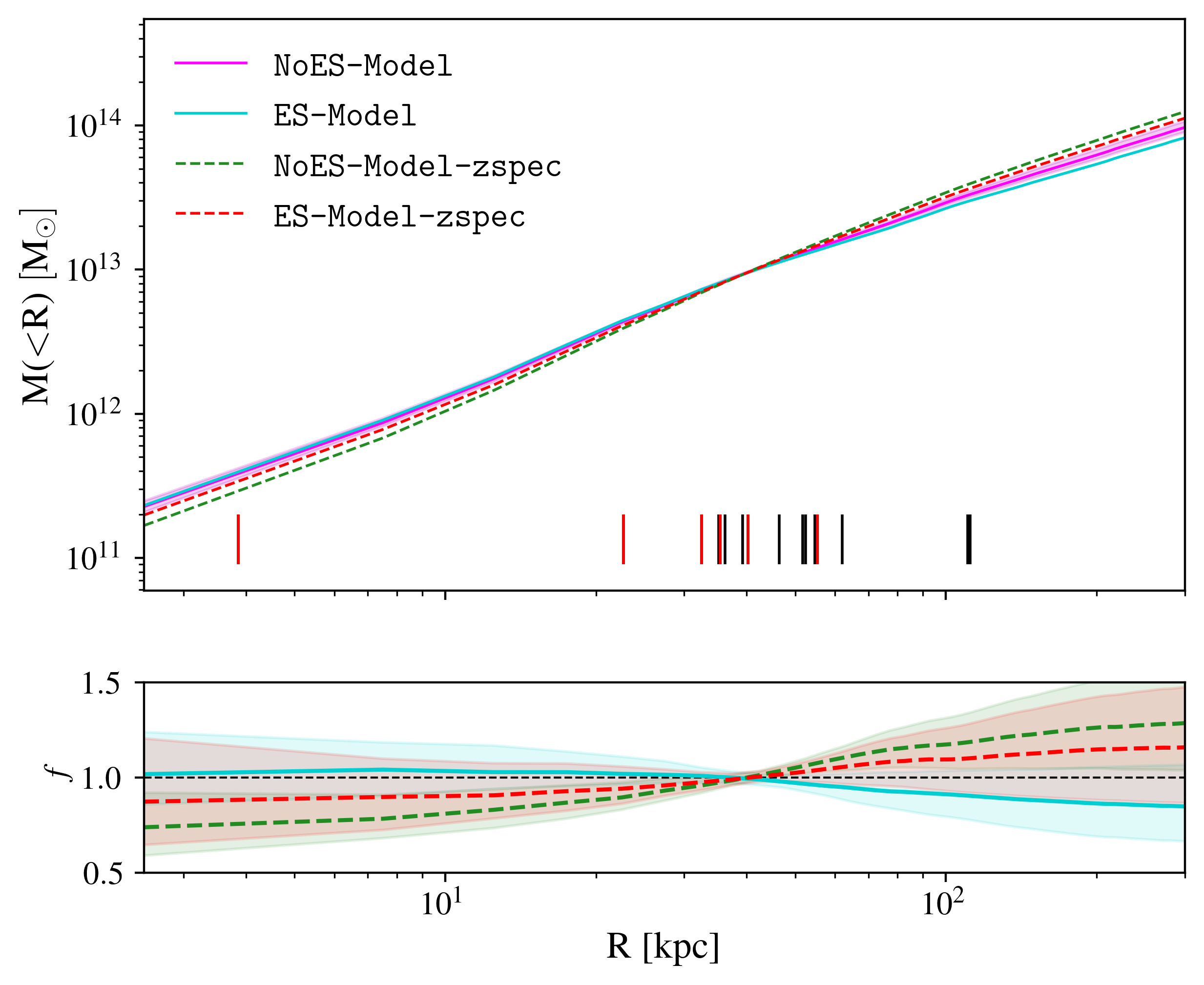} 
\caption{Cumulative projected total mass profile of SDSS 2222 as a function of the distance from the BCG centre (ID 2010, marked as a black cross in Fig. \ref{Fig:multimg}). 
\textit{Top:} The coloured lines show the median values of the total mass profile, and the shaded magenta area encompasses the 16th and the 84th percentiles, estimated from 500 random Bayesian Markov chain Monte Carlo realisations for the \texttt{NoES-Model}.
The projected distances of the 18 multiple images from the BCG are indicated as vertical black lines. The distances of the six quasar images are highlighted in red. \textit{Bottom:} Ratio between the projected total mass profiles obtained for the different colour-coded models and the \texttt{NoES-Model}.
\label{fig:massprof}
}
\end{figure}

\begin{table*}
\caption{Median values and 68\% (statistical) confidence level intervals of the lens mass parameters for the SL models discussed in this work.} 
\label{tab:massparams}
\centering
\renewcommand{\arraystretch}{1.2}
\begin{tabular}{c|ccccccccccc}
\hline
Model & Component & $x$ & $y$ & $e$ & $\theta$ & $\sigma_0$ & r$_{\rm cut}$ & r$_{\rm core}$ & $\gamma_{\rm ext}$ & $\phi$\\ 
&  & [\arcsec] & [\arcsec] &   & [$\deg$] & [km\ s$^{-1}$] & [\arcsec] & [\arcsec] &  & [$\deg$]\\
\hline
\texttt{NoES-Model} &  DM & $0.0^{+0.2}_{-0.2}$ & $1.8^{+0.1}_{-0.4}$ & $0.58^{+0.06}_{-0.06}$ & $94.9^{+1.0}_{-1.0}$ & $705^{+36}_{-25}$& [2000] & $5.6^{+1.0}_{-0.8}$& -- & --\\ 
&  $L^{\star}$ Galaxy & -- & -- & -- & -- & $345^{+15}_{-13}$ & $3.4^{+0.7}_{-0.5}$ & [0] & -- & --\\
\hline
\texttt{ES-Model} &  DM & $0.0^{+0.2}_{-0.2}$ & $1.7^{+0.1}_{-0.4}$ & $0.46^{+0.11}_{-0.13}$ & $103.2^{+8.9}_{-3.6}$ & $620^{+56}_{-79}$& [2000] & $4.2^{+1.5}_{-1.2}$& -- & --\\ 
&  Ext. Shear & -- & -- & -- & -- & -- & -- & -- & $0.12^{+0.05}_{-0.08}$ & $169.2^{+2.7}_{-26.4}$\\
&  $L^{\star}$ Galaxy & -- & -- & -- & -- & $345^{+16}_{-11}$ & $4.1^{+1.5}_{-0.7}$ & [0] & -- & --\\

\hline
\texttt{NoES-Model-zspec} &  DM & $0.0^{+0.1}_{-0.1}$ & $1.1^{+0.6}_{-0.5}$ & $0.35^{+0.09}_{-0.08}$ & $92.9^{+0.9}_{-0.8}$ & $805^{+68}_{-39}$& [2000] & $6.3^{+1.6}_{-1.4}$& -- & --\\ 
&  $L^{\star}$ Galaxy & -- & -- & -- & -- & $288^{+24}_{-23}$ & $2.1^{+0.9}_{-0.8}$ & [0] & -- & --\\
\hline
\texttt{ES-Model-zspec} &  DM & $-0.2^{+0.2}_{-0.3}$ & $1.3^{+0.1}_{-0.9}$ & $0.47^{+0.15}_{-0.15}$ & $98.6^{+3.9}_{-5.2}$ & $756^{+67}_{-56}$& [2000] & $5.4^{+1.7}_{-1.4}$& -- & --\\ 
&  Ext. Shear & -- & -- & -- & -- & -- & -- & -- & $0.07^{+0.06}_{-0.04}$ & $134.1^{+17.1}_{-38.9}$\\
&  $L^{\star}$ Galaxy & -- & -- & -- & -- & $326^{+36}_{-31}$ & $2.1^{+0.8}_{-0.7}$ & [0] & -- & --\\
\hline
\end{tabular}
\tablefoot{
Parameter values in square brackets are kept fixed in the optimisation. We note that the $L^{\star}$ corresponds to the reference luminosity adopted to be that of the BCG with a magnitude value in the HST F160W band of $17.56$.
 }
\end{table*}

\subsection{Mass parametrisation of SDSS 2222} \label{sec:MassModel}
In this work we consider two sets of cluster total mass parametrisations and two sets of observables in order to find which model best fits the data and to investigate the impact of systematic uncertainties on our results.

\noindent \textit{Mass models:}
The cluster-scale mass component of SDSS 2222 was modelled with a single, non-truncated dPIE profile. The values of the ellipticity, position angle, core radius, and velocity dispersion associated with this halo were optimised within large flat priors, while the truncation radius was fixed to a very large value. The coordinate values of the halo were free to vary within $2\arcsec$ of those of the BCG. 
For the galaxy-scale mass component, we considered the 37 spectroscopically confirmed cluster members (see Sect. \ref{sec:cm} and Table \ref{tab:cm}), which were all modelled within the adopted scaling relations (see Eq. \ref{eqsigma}). As presented in Sect. \ref{sec:cm}, we made use of the stellar velocity dispersion measurements obtained for a sub-set of cluster galaxies. In particular, in the lens model optimisations we used the normalisation and slope values of the best-fit $\sigma_0$--F160W relation. We  thus chose a Gaussian distribution centred on the measured value of $321~\rm km~s^{-1}$ and with a standard deviation value of $41 \mathrm{\, km \, s^{-1}}$ as a prior for the value of the normalisation $\sigma_{0}^{\star}$. Instead, since no independent information is available, a large flat prior for the $r_{\rm cut}^{\star}$ value was considered (between 0 and 250 kpc). The value of the slope, $\alpha$, was fixed to the fitted one of 0.295 (based on the stellar kinematic measurements; see Sect. \ref{sec:cm}), and the value of $\beta$ was inferred such that $M^{\rm tot}_i/L_i \propto L ^{0.2}_i$ following
\begin{equation}
\label{eqbeta}
\beta=\gamma-2\alpha+1.
\end{equation}
This cluster total mass parametrisation is referred to as \texttt{NoES-Model}.
In the second mass model, labelled \texttt{ES-Model}, we included an additional external shear component, as described in Sect. \ref{sec:method}.
\texttt{NoES-Model} has a total of 8 free parameters related to the mass parametrisation, while \texttt{ES-Model} has 10.\\

\noindent \textit{Observables:}
The full catalogue of multiple images is presented in Sect. \ref{sec:multim} and Table \ref{tab:multim}. In total, we have five multiply imaged sources, three of which are spectroscopically confirmed.
To investigate potential biases coming from the inclusion of less secure information \citep[see e.g.][]{Grillo2015, Johnson2016}, we considered the two lens total mass models presented above, including either the full spectroscopic and photometric multiple image sample or the spectroscopic-only one (labelled \texttt{NoES-Model-zspec} and \texttt{ES-Model-zspec}). In the former case, the redshift values of systems 3 and 4 were optimised in the models, assuming large flat priors.
These catalogues thus provide 36 or 24 observational constraints, and the lens models depend on 12 or 6 free parameters for the positions and redshifts of the corresponding sources, respectively.\\

Initially, a positional uncertainty of $0\arcsec.25$ was adopted for all images and models. For each final model run, the multiple image positional uncertainty was then re-scaled in order to obtain a minimum $\chi^2$ value comparable with the number of degrees of freedom ($\nu$) such that  $\chi^2/\nu\sim1$. In particular, we re-scaled the positional uncertainty for both \texttt{NoES-Model} and \texttt{ES-Model} from $0\arcsec.25$ to $0\arcsec.35$, while for \texttt{NoES-Model-zspec} and \texttt{ES-Model-zspec} we used the value of $\chi^2/\nu\sim1$ without any re-scaling of the positional uncertainty (see Table \ref{tab:FoM}).

\section{Results and discussion} \label{sec:results}
In this section we present the results from the model optimisations and statistical analyses described in Sect. \ref{sec:MassModel} and discuss the impact on our results of systematic uncertainties related to the modelling assumptions and choices of observables.

We refer to two cluster total mass parametrisations (with or without an external shear component) and two sets of observational constraints (a full or spectroscopic-only multiple image sample).
The values of the statistical estimators introduced in Sect.~\ref{sec:method} for these four SL models are summarised in Table \ref{tab:FoM}, and the resulting median values of the model free parameters, and the associated $1\sigma$ statistical uncertainties, are given in Table \ref{tab:massparams}. 
Considering the values of all figures of merit, \texttt{ES-Model} was favoured when considering the full sample of multiple images, despite having two additional free parameters related to the external shear field. However, when considering only the spectroscopic sample of lensed sources, the inclusion of an external shear component is not supported, given the larger BIC and AIC values with respect to the model without. 
The resulting rms value for the  best-fit model with the full sample of images (i.e. \texttt{ES-Model}) is $0\arcsec.29$.

Figure \ref{Fig:multimg} shows the resulting critical lines at the redshift of the QSO system ($z=2.801$) for the four models. The main difference arises between models that consider different sets of observables (see the solid and dashed coloured lines), in the southern region of the cluster core, and they can be attributed to the degeneracy between the cluster-scale and galaxy-scale mass components. 
As previously mentioned, including as additional information the surface brightness distribution of the southern arc, A1, should help robustly reconstruct the total mass distribution of the lens and distinguish between different total mass parameterisations.

The obtained values of the model mass parameters are generally consistent, within the statistical uncertainties. In addition, Fig. \ref{Fig:sigma} highlights the importance of an independent determination and implementation in our lens models of the $\sigma_0$--F160W sub-halo scaling relation in reducing inherent model degeneracies between the cluster- and galaxy-scale mass components.
When contrasted with the previous SL analysis of SDSS 2222 by \citetalias{Sharon2017}, the values of the mass parameters related to the large-scale dark matter (DM) halo are consistent, within the errors. However, a comparison of the sub-halo contribution to the total mass is not straightforward: in \citetalias{Sharon2017}, the value of the normalisation of the velocity dispersion--luminosity scaling relation was fixed to a given arbitrary value and the value of the reference luminosity, $L^{\star}$, was not provided.

We find that when an external shear component is included, the \texttt{ES-Model}(\texttt{-zspec}) models require a non-negligible amplitude (i.e. $\gamma_{\rm ext}=0.12(0.07))$, albeit with a large statistical uncertainty. Similar values have been quoted in the literature \citep[see e.g.][]{Caminha2016, Caminha2019, Lagattuta2019}. The inclusion of this additional (non-localised) term introduces further flexibility into the lens modelling and can account for several non-modelled lensing effects, such as the cluster environment, line-of-sight mass structures, or asymmetries in the total mass distribution \citep[see e.g.][and \citetalias{Acebron2022}, for a discussion]{Lagattuta2019}.
We also find consistent model predictions for the redshift value of system 3: $z_{\rm S3}=3.3_{-0.2}^{+0.4}$ for \texttt{NoES-Model} and $z_{\rm S3}=3.2_{-0.2}^{+0.3}$ for \texttt{ES-Model}. These values are also in agreement with the tentative GMOS spectroscopic measurement and the lens model predictions from \citetalias{Sharon2017}.
On the other hand, the model-predicted redshift value for system 4 is completely unconstrained in both lens models. System 4 being the farthest identified system from the cluster centre, but angularly close to the third brightest cluster member ($\sim3\arcsec$ from galaxy ID 2064 in Table \ref{tab:cm}), this can be explained by model degeneracies between the system redshift and the sub-halo mass component.


\begin{figure*}
\centering
\includegraphics[width=\columnwidth,trim=170mm 53mm 170mm 0mm, clip=true]{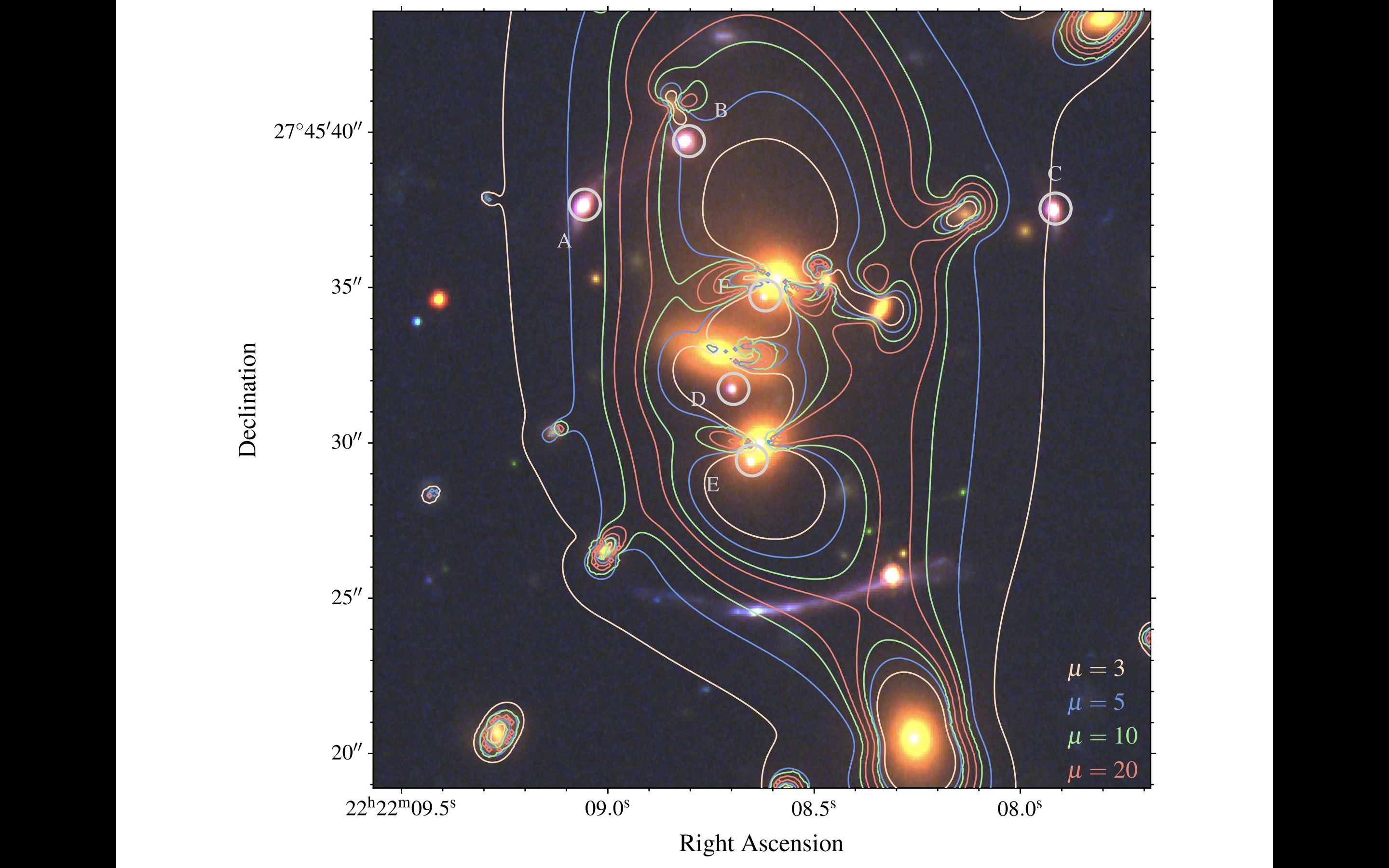}
\includegraphics[width=0.881\columnwidth,trim=250mm 53mm 170mm 0mm, clip=true]{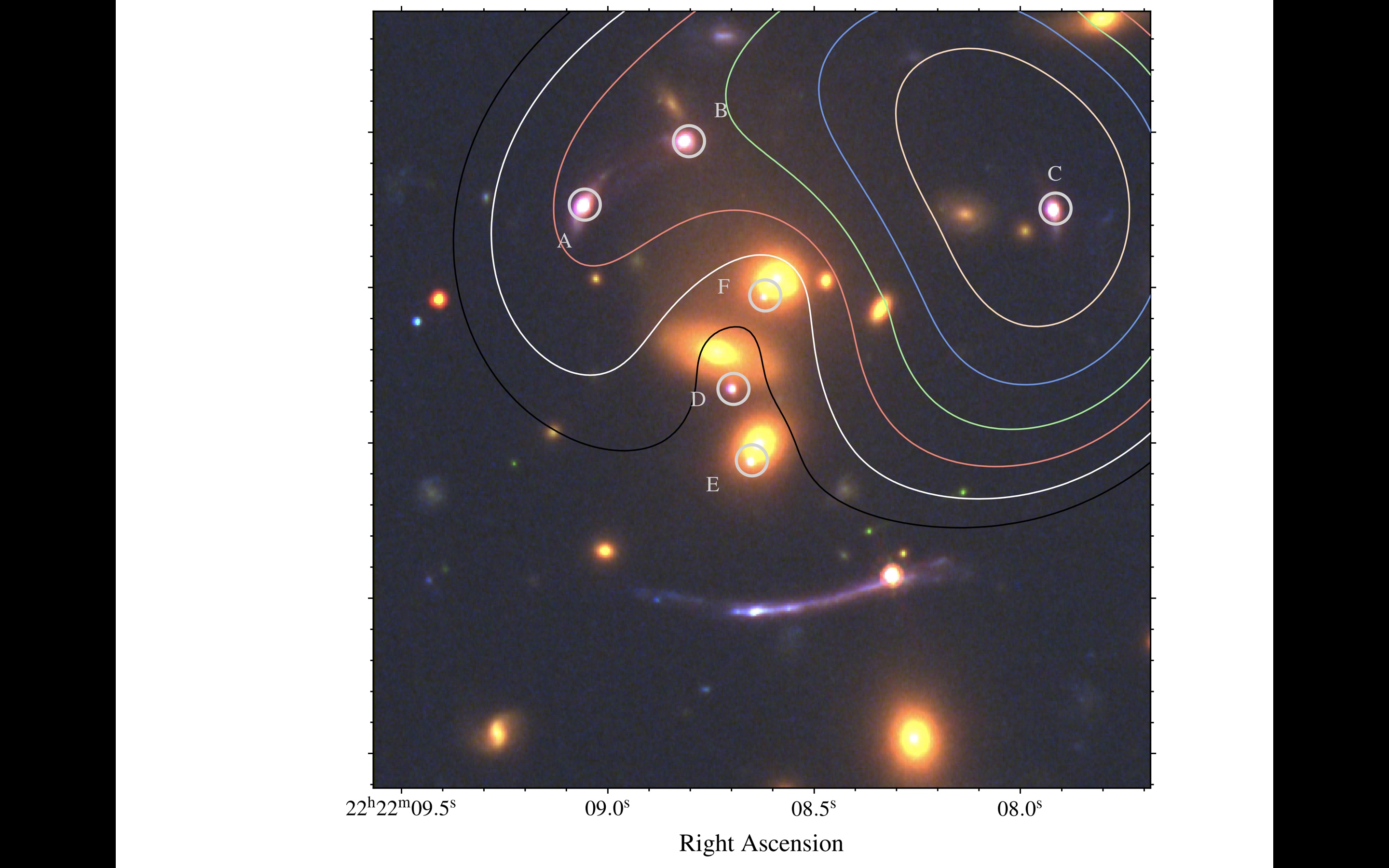}\\
\includegraphics[width=\columnwidth,trim=170mm 8mm 170mm 0mm, clip=true]{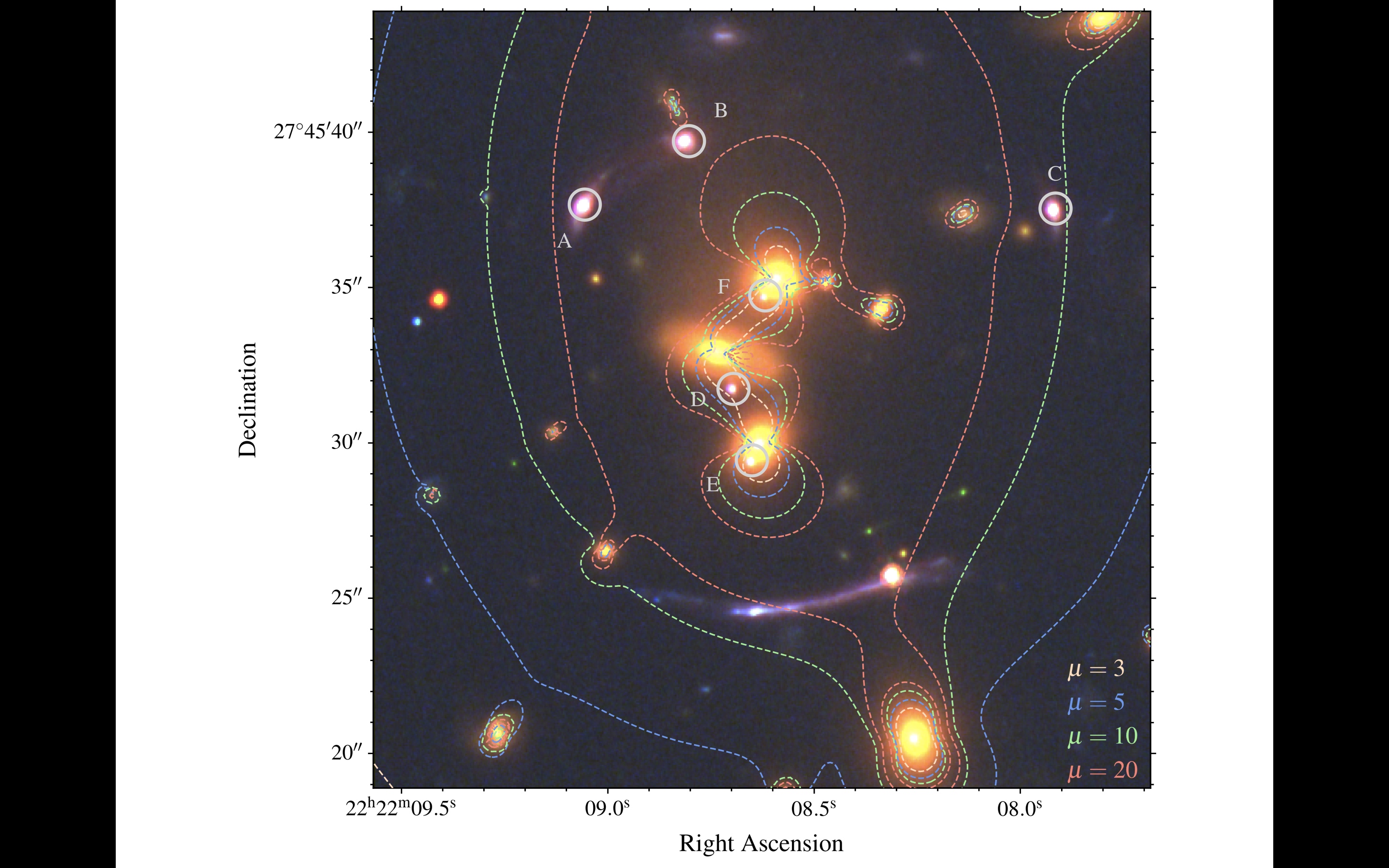}
\includegraphics[width=0.881\columnwidth,trim=250mm 8mm 170mm 0mm, clip=true]{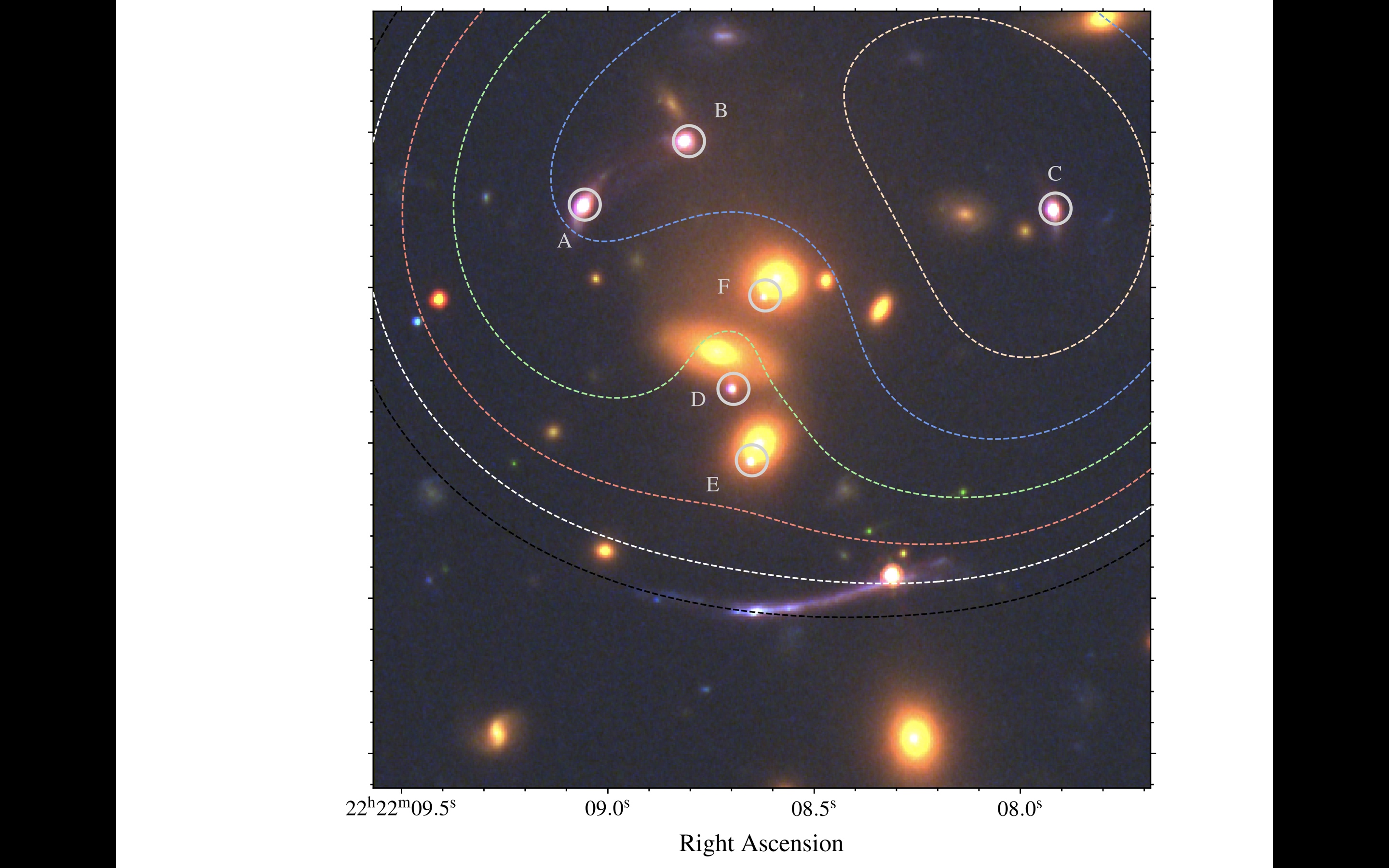}
\caption{Absolute magnification (\textit{left}) and linearly spaced Fermat potential (\textit{right}) contour levels for a source at the redshift of the QSO system, $z = 2.801$. The \textit{top} and \textit{bottom} panels correspond to the best-fit \texttt{ES-Model} and \texttt{NoES-Model-zspec}, respectively. In all panels, the positions of the QSO images are indicated as light grey circles.
\label{fig:magzqso}}
\end{figure*}

We show in Fig. \ref{fig:massprof} the cumulative projected total mass profile of the cluster as a function of the distance from the BCG centre for the four lens models (top panel) and the ratio of the profiles from the different models with respect to \texttt{NoES-Model} (bottom panel). 
Within the approximate average distance of the multiple images from the BCG centre, we measure a precise projected total mass value of $M(<40~\rm kpc)= 1.02_{-0.02}^{+0.02} \times 10^{13}~ M_{\odot}$ for \texttt{ES-Model}. Given the errors, this mass estimate is perfectly in agreement for the four models, with a statistical plus systematic relative uncertainty (determined from the different mass parametrisations) of only approximately $3\%$.
The projected total mass of the cluster enclosed within a circle with a radius equal to the average multiple image distance from the lens centre (indicated as vertical lines in Fig. \ref{fig:massprof}) is thus robustly measured. Well beyond the region where the observables are available, model extrapolations and systematic uncertainties become more relevant, but all model projected total mass profiles remain consistent, within $2\sigma$.

SDSS 2222 is a relatively low-mass lens cluster, with an X-ray luminosity in the group regime, a complex morphology, and a fairly small number of securely identified {`point-like'} multiple image constraints. We note that current SL models suffer from important degeneracies between several model parameters. In particular, we verified that the $y$ coordinate value of the main DM halo is correlated to those of its velocity dispersion and ellipticity, and the mass contribution of the sub-halo component to the lens total mass. We thus further investigated the impact of relaxing or tightening the prior on the $y$ coordinate value of the main DM halo (from $1\arcsec$ to $10\arcsec$ from the BCG). We find that models for which a larger $y$ value is favoured (i.e. with a cluster DM halo more distant from the luminous counterpart) require a rounder (i.e. less elliptical) DM halo, with a more massive sub-halo component.
Despite these degeneracies (as illustrated in Fig. \ref{fig:posterior} for \texttt{ES-Model} and \texttt{NoES-Model-zspec}), the different lens models yield projected total mass profiles, normalisations of the $\sigma_0$--F160W sub-halo scaling relation, and critical lines (e.g. at the redshift of the QSO) that are consistent within the statistical uncertainties, showing that these global quantities and features are robustly reconstructed.

However, we remark that the model-predicted magnification and time-delay values are much more sensitive to the modelling assumptions and considered constraints, and thus to systematic uncertainties. Magnifications and time delays of the multiple images of a source do indeed depend on the local details, close to the multiple-image-observed positions, of a lens total mass density distribution.
This is illustrated in Fig. \ref{fig:magzqso}, where we compare the magnification and Fermat potential (or the arrival time-delay surface) contours for a source at the position and redshift of the QSO, obtained from the two best-fit \texttt{ES-Model} and \texttt{NoES-Model-zspec} models. Clearly, they provide significantly different predictions.
In particular, we find that the model-predicted time delays between the QSO multiple image pairs A and B and A and C, which have measured time delays \citep{Dahle2015, Dyrland2019}, can vary by factors of approximately 1.4 and 1.5, respectively. When different priors on the $y$ coordinate value of the DM cluster-scale halo are adopted, we obtain similar variations for the model-predicted time delays, showing how sensitive these quantities are to the modelling details. Therefore, we stress that if one is interested in magnifications and time delays, it is not possible to distinguish among the different predictions of disparate models that can reproduce the observed positions of a set of {point-like} multiple images similarly well.
We thus report similar findings to those presented in \citetalias{Acebron2022}, in the sense that cluster SL models with comparable rms and statistical estimator values, referring to only the positions of multiple images, can provide contrasting values of predicted time delays, which would then result in considerably different estimates of the Hubble constant. These results further highlight the need to include the measured values and associated uncertainties of the time delays between the multiple images of a variable source as observational constraints for cluster lensing cosmological applications \citep[see e.g.][]{Grillo2018, Grillo2020}.

\section{Conclusions} \label{sec:conclusions}
SDSS J2222+2745, at $z=0.489$, is one of the few currently known lens clusters that host multiple  images (six)  of a background quasar ($z=2.801$) with measured time delays between two image pairs \citep{Dahle2015, Dyrland2019}. In order to exploit this particular lens cluster as a robust cosmological probe, a high-precision and accurate lens model is crucial.
As a first step towards future cosmological analyses, in this work we have presented and used recent VLT/MUSE spectroscopic observations of the SDSS J2222+2745 core. In combination with archival multi-band HST imaging, we have been able to securely identify more than 30 cluster members and confirm the redshift values of three multiple image families. Based on these new data, we have built a refined SL model of the galaxy cluster SDSS J2222+2745 with the parametric software \texttt{lenstool} \citep{Jullo2007}.

Our findings can be summarised as follows: 

\begin{enumerate}
    \item Thanks to the MUSE spectra, we have provided an updated redshift value for the lensed quasar. In addition, we have measured secure redshifts for all images in system 2, for which only one of the three  images had previously been spectroscopically confirmed (\citetalias{Sharon2017}).
    Our SL models have included five families (three of them spectroscopic), with a total of 18 multiple images (see Table \ref{tab:multim} and Figs. \ref{Fig:multimg} and \ref{fig:spec}). \\
    \item We have spectroscopically identified 34 cluster galaxies with a QF $\geq$ 2 (see Table \ref{tab:cm} and Fig. \ref{Fig:MUSE}), a pure sample that was then included in the SL modelling. By further exploiting the new MUSE data, we have reliably measured the stellar velocity dispersion values of a sub-sample of 13 cluster members, down to HST F160W $\sim$ 21. These values have been used to independently calibrate the scaling relations of the sub-halo population in the lens modelling (see Fig. \ref{Fig:sigma}). \\
    \item We have performed a parametric SL modelling of SDSS J2222+2745. The observed positions of the multiple images are reproduced with an rms value of $0\arcsec.29$ for the best-fit \texttt{ES-Model} model (see Table~\ref{tab:FoM}). Within the average projected distance of the multiple images from the BCG centre, we have measured a precise cluster total mass value of $M(<40~\rm kpc)= 1.02 \times 10^{13}~ M_{\odot}$, with a statistical and systematic relative uncertainty of approximately $3\%$. \\ 
    \item We have investigated the impact of systematic uncertainties arising from different mass parametrisations, sets of observational constraints, and model degeneracies. We find that the lens projected total mass profile, the normalisation of the $\sigma_0$--F160W sub-halo scaling relation, and the critical lines (at the redshift of the quasar) are all robustly measured. \\
    \item We have verified that the model-predicted magnification and time-delay values are very sensitive to the reconstructed local distribution of the lens total mass density, and thus to systematic uncertainties. 
    In particular, we have shown, from similar lens total mass models and sets of observational constraints, that the predicted time delays between the quasar multiple image pairs A and B and A and C (both with measured time delays) can vary by more than 30\%. This finding further stresses the fact that time-delay predictions obtained from different {point-like} lens models that do not include the measured time delays as constraints should not be used to estimate the value of the Hubble constant.

\end{enumerate}
SDSS J2222+2745 is a complex cluster with a relatively small number of secure multiply imaged sources. 
Going beyond {point-like} SL models, by including the surface brightness distribution of the multiple images of the quasar host galaxy and of the southern tangential arc, will alleviate current model degeneracies and result in a significant improvement of the modelling robustness. This will in turn provide more accurate and precise lens total mass reconstructions and predictions for the multiple image systems. 
As already remarked in \citetalias[]{Acebron2022}, to fully exploit cluster-scale lensing systems such as SDSS J2222+2745 as cosmological probes, it is key that all available lensing observables, including the measured values and errors of the time delays \citep{Grillo2018, Grillo2020}, be incorporated in the analyses.
The full MUSE spectroscopic catalogue of SDSS J2222+2745 presented in this work is made publicly available\footnote{The catalogue is available at \url{www.fe.infn.it/astro/lensing}.}.

\begin{acknowledgements}
We kindly thank the anonymous referee for the useful comments that have helped improving the manuscript.
This work is based in large part on data collected at ESO VLT (prog. ID 0103.A-0554(A)) and NASA \textit{HST}.
AA has received funding from the European Union’s Horizon 2020 research and innovation programme under the Marie Skłodowska-Curie grant agreement No 101024195 — ROSEAU.
We acknowledge financial support through grants PRIN-MIUR 2017WSCC32 and 2020SKSTHZ.
MN aknowledges INAF Mainstream 1.05.01.86.20.
GBC thanks the Max Planck Society for support through the Max Planck Research Group for S. H. Suyu and the academic support from the German Centre for Cosmological Lensing.
\end{acknowledgements}

%
%

\bibliographystyle{aa}
\bibliography{bibliography}


\begin{appendix}

\section{Multiple images in SDSS 2222}
\begin{figure*}[hbt!]
\centering
\includegraphics[width=\linewidth]{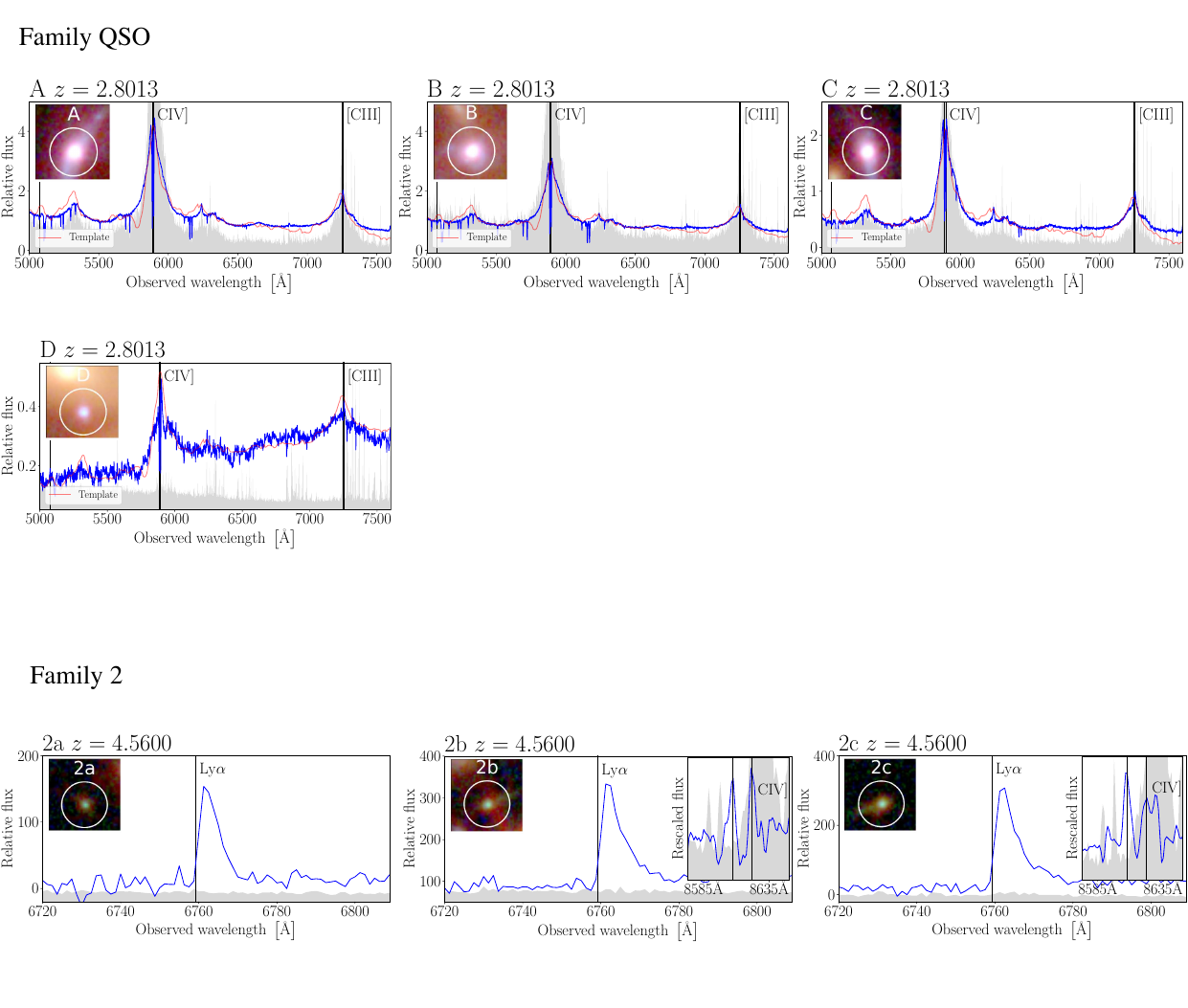} 
\caption{MUSE data of the multiply imaged background sources identified in SDSS 2222 with QF $\ge$ 2 (see Table \ref{tab:multim}). The vertical black lines indicate the positions of the emission lines based on the best estimate of the systemic redshift. The grey area shows the re-scaled variance obtained from the data reduction pipeline. The flux is given in units of 10 $\rm erg~s^{-1}~cm^{-2}$ \r{A}$^{-1}$. The image cutouts in each panel, which are  $2\arcsec$ across, are extracted from the colour-composite HST image; the white circles pinpoint the HST counterparts.} \label{fig:spec}
\end{figure*}

We present in Fig. \ref{fig:spec} the MUSE spectra of the multiply imaged background sources securely identified in SDSS 2222, together with the colour-composite HST counterparts.

\clearpage

\section{Cluster members included in the SL modelling of SDSS 2222}
We present in Table \ref{tab:cm} the coordinates, spectroscopic redshifts, and the associated QF for the cluster galaxies that are considered in the SL model. 

\begin{table}[h]
\caption{Catalogue of the spectroscopic cluster members included in the SL modelling of SDSS 2222.
} 
\label{tab:cm}
\centering
\begin{tabular}{ccccc}
\hline\hline
ID & R.A. & Decl & $\rm z_{spec}$ & QF\\ 
 & deg & deg &  &  \\ 
\hline
2010 & 335.535746 & 27.759866 & 0.4900 & 3\\ 
1974\tablefootmark{a} & 335.546683 & 27.758010 & 0.4833 & -\\ 
2064 & 335.534337 & 27.755753 & 0.4917 & 3\\ 
1682 & 335.536303 & 27.759224 & 0.4926 & 3\\ 
1351\tablefootmark{a} & 335.535539 & 27.772641 & 0.4833 & -\\ 
1879 & 335.535913 & 27.758409 & 0.4910 & 3\\ 
2514 & 335.527403 & 27.751236 & 0.4899 & 3\\ 
1614 & 335.532449 & 27.762223 & 0.4873 & 3\\ 
254\tablefootmark{a} & 335.536865 & 27.744699 & 0.4945 & -\\ 
3675 & 335.533015 & 27.765765 & 0.4899 & 3\\ 
2392 & 335.533646 & 27.753336 & 0.4861 & 3\\ 
1859 & 335.534695 & 27.759614 & 0.4977 & 3\\ 
2277 & 335.535656 & 27.755222 & 0.4895 & 3\\ 
1782 & 335.533871 & 27.760461 & 0.4890 & 3\\ 
2175 & 335.538561 & 27.755814 & 0.4857 & 3\\ 
2285 & 335.535389 & 27.754893 & 0.4871 & 3\\ 
1676 & 335.535248 & 27.759868 & 0.4904 & 3\\ 
2459 & 335.534338 & 27.752224 & 0.4839 & 3\\ 
2129 & 335.531878 & 27.756635 & 0.4834 & 3\\ 
2051 & 335.537469 & 27.757450 & 0.4927 & 3\\ 
1959 & 335.536778 & 27.761426 & 0.4872 & 3\\ 
3603 & 335.533086 & 27.767028 & 0.4882 & 3\\ 
2523 & 335.532022 & 27.751610 & 0.4914 & 3\\ 
3201 & 335.540378 & 27.765003 & 0.4906 & 3\\ 
2021 & 335.529668 & 27.757910 & 0.4928 & 3\\ 
1791 & 335.528265 & 27.760577 & 0.4935 & 3\\ 
1835 & 335.530911 & 27.759113 & 0.4894 & 3\\ 
232 & 335.540238 & 27.766579 & 0.4829 & 3\\ 
328 & 335.530760 & 27.761539 & 0.4889 & 3\\ 
1631 & 335.537592 & 27.762424 & 0.4800 & 3\\ 
1993 & 335.537994 & 27.758516 & 0.4862 & 3\\ 
423 & 335.528033 & 27.756864 & 0.4967 & 3\\ 
294 & 335.533328 & 27.762795 & 0.4906 & 3\\ 
2022 & 335.539222 & 27.757964 & 0.4984 & 3\\ 
2407 & 335.529515 & 27.753380 & 0.4898 & 2\\ 
265 & 335.543839 & 27.764291 & 0.4896 & 2\\ 
1801 & 335.538663 & 27.760608 & 0.4803 & 3\\ 
\hline  
\end{tabular}
\tablefoot{
\tablefoottext{a}{Redshift values from the SDSS DR9 catalogue.}\\
}
\end{table}

\clearpage

\section{Foreground and background galaxies identified in SDSS 2222}
We present in Table \ref{tab:fgbg} the coordinates and spectroscopic redshifts, with the associated QF, of foreground ($z<0.474$) and background ($z>0.504$) galaxies with respect to the cluster redshift.

\begin{table}[h]
\caption{Catalogue of the foreground (top) and background (bottom) galaxies with a secure spectroscopic measurement based on the MUSE data.
} 
\label{tab:fgbg}
\centering
\begin{tabular}{cccccc}
\hline\hline
ID & R.A. & Decl & $\rm z_{spec}$ & QF\\ 
 & deg & deg &   &  \\ 
\hline
\vspace{-0.2cm}\\
2494 & 335.527883 & 27.752258 & 0.1666 & 3\\ 
2060 & 335.535654 & 27.756604 & 0.1731 & 3\\ 
1836 & 335.531605 & 27.759843 & 0.2844 & 3\\ 
1566 & 335.533400 & 27.763878 & 0.2846 & 3\\ 
2161 & 335.531964 & 27.759812 & 0.2852 & 3\\ 
2219 & 335.545558 & 27.755106 & 0.4657 & 3\\ 
2218 & 335.545291 & 27.754859 & 0.4659 & 3\\ 
\hline  
2520 & 335.546073 & 27.751610 & 0.603 & 3\\ 
1787 & 335.539611 & 27.760557 & 0.624 & 3\\ 
341 & 335.539542 & 27.760744 & 0.624 & 3\\ 
2455 & 335.529934 & 27.752699 & 0.666 & 3\\ 
2167 & 335.532660 & 27.756038 & 0.686 & 3\\ 
555556 & 335.534018 & 27.755143 & 0.686 & 3\\ 
1579 & 335.533869 & 27.762893 & 0.687 & 3\\ 
1565 & 335.540698 & 27.763180 & 0.718 & 3\\ 
482 & 335.542613 & 27.754734 & 0.754 & 3\\ 
2196 & 335.541723 & 27.755398 & 0.832 & 3\\ 
1673 & 335.535118 & 27.757902 & 0.832 & 3\\ 
1733 & 335.545648 & 27.761332 & 0.834 & 2\\ 
3021 & 335.526249 & 27.751953 & 0.853 & 3\\ 
2217 & 335.544759 & 27.754999 & 0.908 & 3\\ 
462 & 335.544976 & 27.755142 & 0.909 & 3\\ 
3464 & 335.527385 & 27.767330 & 0.910 & 3\\ 
297 & 335.528105 & 27.762531 & 0.981 & 3\\ 
2598 & 335.535789 & 27.750941 & 1.024 & 2\\ 
2382 & 335.531780 & 27.753580 & 1.070 & 3\\ 
480  & 335.528616 & 27.754721 & 1.071 & 3\\ 
2401 & 335.545781 & 27.752973 & 1.173 & 2\\ 
1696 & 335.534418 & 27.761781 & 1.200 & 3\\ 
1753 & 335.536373 & 27.761947 & 1.201 & 3\\ 
55555151 & 335.530780 & 27.765559 & 1.269 & 9\\
2300 & 335.539642 & 27.754210 & 1.272 & 3\\ 
513 & 335.532463 & 27.753210 & 1.273 & 3\\ 
\hline  
\end{tabular}
\end{table}

\begin{table}[h]
\vspace{4.05cm}
\centering
\begin{tabular}{cccccc}
\hline\hline
ID & R.A. & Decl & $\rm z_{spec}$ & QF \\ 
 & deg & deg &  &   \\ 
\hline
\vspace{-0.2cm}\\
1461 & 335.538489 & 27.764606 & 1.295 & 3\\ 
387 & 335.530384 & 27.758833 & 1.314 & 2\\ 
1927 & 335.544663 & 27.759414 & 1.510 & 2\\ 
1731 & 335.528905 & 27.761430 & 1.536 & 2\\ 
3557 & 335.533823 & 27.767280 & 2.039 & 2\\ 
2261 & 335.534169 & 27.757256 & 2.176 & 2\\ 
2150 & 335.536000 & 27.756825 & 2.295 & 3\\ 
2061 & 335.536858 & 27.756943 & 2.295 & 3\\ 
2134 & 335.535668 & 27.756852 & 2.296 & 3\\ 
555559 & 335.529544 & 27.757045 & 3.060 & 3\\ 
533 & 335.542052 & 27.752321 & 3.131 & 3\\ 
999991 & 335.532275 & 27.757333 & 3.277 & 9\\ 
55555141 & 335.536801 & 27.765589 & 3.280 & 9\\ 
528 & 335.529304 & 27.752584 & 3.454 & 9\\ 
5555514 & 335.527340 & 27.761884 & 3.495 & 3\\ 
5555513 & 335.527428 & 27.762041 & 3.495 & 9\\ 
999993 & 335.545637 & 27.765787 & 3.680 & 3\\ 
2091 & 335.539297 & 27.757099 & 3.869 & 9\\ 
3642 & 335.538114 & 27.766197 & 3.870 & 9\\ 
55555154 & 335.529808 & 27.764538 & 3.909 & 3\\ 
555557 & 335.537407 & 27.754958 & 4.531 & 3\\ 
555552 & 335.538607 & 27.754902 & 4.538 & 3\\ 
55555115 & 335.533177 & 27.755314 & 4.538 & 2\\ 
479 & 335.538535 & 27.754866 & 4.540 & 3\\ 
478 & 335.538719 & 27.754932 & 4.542 & 2\\ 
488 & 335.536164 & 27.754437 & 4.548 & 2\\ 
308 & 335.527973 & 27.762278 & 4.559 & 2\\ 
472 & 335.542556 & 27.755287 & 4.562 & 3\\ 
55555113 & 335.529161 & 27.754992 & 4.710 & 3\\ 
5555514 & 335.539716 & 27.766822 & 4.711 & 3\\ 
999994 & 335.543832 & 27.751557 & 4.725 & 9\\ 
55555155 & 335.534001 & 27.763796 & 5.185 & 9\\ 
999995 & 335.527068 & 27.753547 & 6.274 & 9\\
\hline  
\end{tabular}
\end{table}

\clearpage

\section{Posterior probability distributions}
\begin{figure*}[hbt!]
\centering
\includegraphics[width=\linewidth]{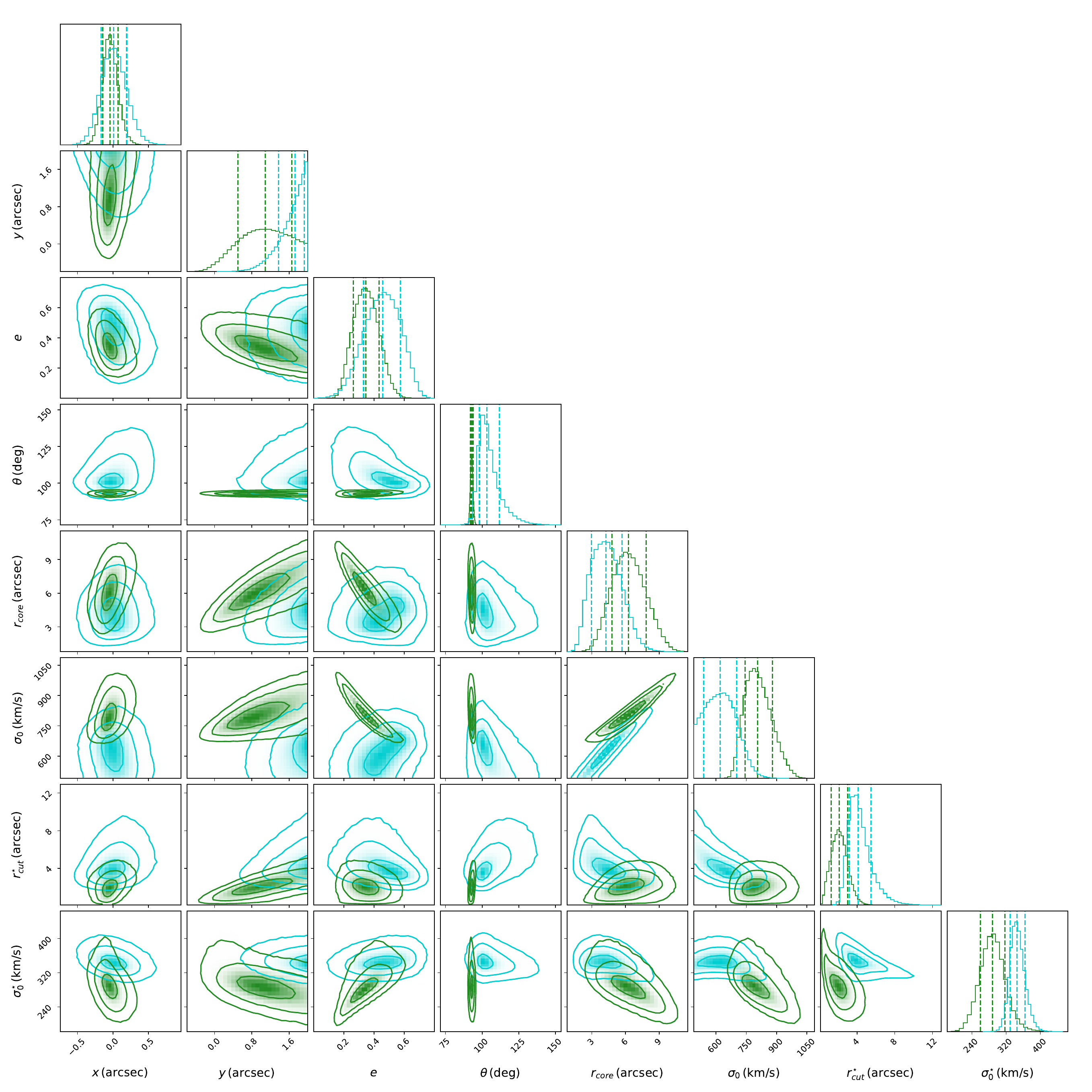} 
\caption{Posterior probability distributions of the parameter values of the cluster-scale and sub-halo mass components for \texttt{ES-Model} (cyan) and \texttt{NoES-Model-zspec} (green). The contours correspond to the 1, 2, and 3$\sigma$ confidence levels, and the vertical dashed lines in the histograms correspond to the 16th, 50th, and 84th percentiles.} \label{fig:posterior}
\end{figure*}

We present in Fig. \ref{fig:posterior} the posterior probability distributions of the parameter values of the cluster-scale and sub-halo mass components for \texttt{ES-Model} and \texttt{NoES-Model-zspec}.

\end{appendix}


\end{document}